\documentclass[journal]{IEEEtran}

\usepackage{threeparttable}
\usepackage{graphicx}
\usepackage{epstopdf}
\usepackage{mathrsfs}
\usepackage{amsmath}
\usepackage{algorithm}
\usepackage{algorithmicx}
\usepackage{algpseudocode}
\usepackage{multirow}
\usepackage{booktabs}
\usepackage[numbers,sort&compress]{natbib}
\usepackage{ntheorem}
\usepackage{stfloats}
\usepackage{bm}
\usepackage{diagbox}
\usepackage{subfigure}
\usepackage{float}
\usepackage{epstopdf}
\usepackage[marginal]{footmisc}
\usepackage{bbding}
\usepackage{pifont}
\usepackage{wasysym}
\usepackage{amssymb}
\usepackage{color}
\usepackage{textcomp}
\usepackage{verbatimbox}
\theoremheaderfont{\upshape}
\theorembodyfont{\upshape}
\theoremseparator{:}

\begin{document}
\title{Deep Learning based Cross-Receiver Radio Frequency Fingerprint Identification \\Under Varying Channels}

\author{Jiashuo He,~\IEEEmembership{Graduate~Student~Member},\IEEEmembership{~IEEE},
        Yumeng Wang,
        Feiyang He,   
        Sai Huang,~\IEEEmembership{Senior~Member},\IEEEmembership{~IEEE},     
        Yiheng Liu,
        Shuo Chang,~\IEEEmembership{Member},\IEEEmembership{~IEEE},  
        and~Zhiyong Feng,~\IEEEmembership{Senior~Member},\IEEEmembership{~IEEE}

\thanks{This work was supported in part by the National Natural Science Foundation of China under Grant (62171045, 62201090, 62321001), and in part by Fundamental Research Funds for the Central Universities under Grant 2023RC16, and in part by the BUPT Excellent Ph.D. Students Foundation under Grant CX2023233.}
\thanks{Jiashuo He, Yumeng Wang, Sai Huang, Yiheng Liu, Shuo Chang, and Zhiyong Feng are with the Key Laboratory of
    Universal Wireless Communications, Ministry of Education, Beijing University
    of Posts and Telecommunications, Beijing 100876, China.
    E-mail: \{jiashuohe, wangyumeng, huangsai, liuyiheng, changshuo, fengzy\}@bupt.edu.cn.
(Corresponding author: Sai Huang.)}
\thanks{Feiyang He is with the Unit 31007 of the PLA, Beijing, 100079, China.
E-mail: flyinghjgc@163.com.}

}

\markboth{}{}

\maketitle

\begin{abstract}
Radio frequency fingerprint identification (RFFI) is a novel paradigm of security mechanism that aims to exploit the hardware characteristics of transmitters for device recognition.
However, the domain variations of receivers (source receiver and target receiver) and wireless channels are key challenges hindering its robustness during the cross-receiver deployment stage.
Despite the fruitful achievements made by many researchers, few works jointly consider the channel and receiver portability issues.
To this end, this paper designs a novel cross-receiver RFFI approach, which also has the merit of channel robustness.
Our training process can be divided into two stages, namely enrollment stage and cross-receiver deployment stage.
In the enrollment stage, we propose a channel-robust signal processing method to construct the denoised spectral quotient (DSQ) sequences.
Then a DSQ-based convolutional neural network (DSQCNN) is designed and trained using the samples collected from the source receiver for transmitter identification.
During the cross-receiver deployment stage, we first employ both the source receiver and target receiver to capture the transmitted signals from a transmitter for building the calibration dataset.
Subsequently, we present a trainable calibration neural network (TCNN) with augmented real-valued input to learn the nonlinear mapping relationship between these two receivers using the calibration dataset.
After accomplishing this, the well-trained TCNN and DSQCNN (TCNN-DSQCNN) will be cascaded with the target receiver to realize the robust cross-receiver classification under varying channel conditions in the classification stage.
As far as we know, this is the first time attempting to solve the channel and receiver portability issues in a joint nonlinearity calibration and channel suppression manner.
To evaluate the performance of the proposed RFFI method, we configure twelve simulated WiFi transmitters and three receivers as a case study.
Simulation results reveal that, for the varying-channel and cross-receiver scenarios, the proposed TCNN-DSQCNN method can achieve robust and reliable classification performance, leading to more than 90$\%$ classification accuracy at SNR = 24 dB.

\end{abstract}


\IEEEpeerreviewmaketitle

\section{Introduction}

\IEEEPARstart{W}{ith} the expeditious advancement in wireless communication technologies, the proliferation of deployed Internet of Things (IoT) devices has undergone a substantial escalation in recent years, with a projected trajectory to reach a quantity of seventy-five billion by 2025 \cite{ref-statistic}.
However, the numerous IoT devices also introduce a pivotal challenge in authentication, necessitating the assurance that received messages are sent from authorized and legitimate devices \cite{ref-authentication}.
Traditional authentication methodologies usually depend on cryptographic solutions, which are susceptible to tampering and demand intensive computational resources \cite{ref-crypt}.
Owing to these constraints, there is a critical need for an effective and resource-efficient way to safeguard IoT devices.

A promising non-cryptographic alternative is known as radio frequency fingerprint identification (RFFI), which exploits the inherent physical-layer hardware impairments of transmitters to extract discriminative information for device authentication, thereby saving computational overhead \cite{ref-overhead}. 
These distinctive hardware impairments typically arise during the manufacturing process due to unintentional operations, which is costly to mimic \cite{ref-RFFs-production}.
Based on that, radio frequency fingerprints (RFFs) resulting from the impairments have been acknowledged as suitable for device recognition.
In this context, the authentication techniques of RFFI have earned escalating interest from both academia and industry.

Currently, RFFI has been widely investigated in both military and civil applications.
In military applications, RFFI is expected to effectively acquire battlefield intelligence regarding the adversary military equipments, thereby gaining information advantages crucial for subsequent strategic adjustments \cite{ref-military}.
In civilian applications, this method can be utilized for the authentication task of wireless communication access control systems \cite{ref-authen}, intrusion detection \cite{ref-iid}, etc \cite{ref-etc1}-\cite{ref-etc2}.
In addition, with the trend of industrial automatic, RFFI can play a vital role in bolstering its security.

\subsection{Motivations}
In wireless communications, the receiver captures the radio frequency (RF) signals transmitted over the air.
It is clear that the received signal will be distorted by both the channel effects and transceiver hardware imperfections.
For a static scenario, the RFFs directly extracted from these RF signals can be effective in identifying transmitters \cite{ref-static}, as the interferences of channel and receiver remain unchanged.
In fact, these fingerprints are intrinsically a kind of coupled RFFs, serving the authentic purpose of discerning the integrated system encompassing the transmitter, channel, and receiver.
Thus, the coupled RFFs are sensitive to any deployment variability arising from channels and receivers.

On the one hand, the real-world RF signals often experience the multipath fading effects attributed to reflections, diffractions, and scattering when encountering obstacles during signal transmission.
Unfortunately, the multipath channels can change dramatically for different propagation environments.
Such uncertainties introduced by variable channels will cause severe distribution shifts in the received RF signals, thereby degrading the transmitters classification performance of the RFFI methods trained in the absence of appropriate countermeasures.
For instance, the authors in \cite{ref-chan-degrade} demonstrated that the classification accuracy can be significantly dropped from 85$\%$ to 9$\%$ when the training and testing data is acquired from different environments.

On the other hand, previous studies have shown that different receivers have distinct effects on the coupled RFFs \cite{ref-diff-receiver1}.
Considering the practical communication networks included massive transceivers, the repeated execution of both data collection and model training for each deployed receiver is ineffective and costly.
Furthermore, even if a single receiver is assumed, it also will be replaced due to malfunctioning or upgrades, thus changing the hardware characteristics in the receiver chain.
As reported from \cite{ref-receiver-degrade}, a small variation in receiver impairments can lead to about 20$\%$ compromise in classification accuracy.

To accurately identify the transmitters, it is reasonable to consider the coupled effects of channels and receivers on the transmitter fingerprints.
Some existing works attempt to solve these problems in RFFI tasks partially \cite{ref-channel-agnostic-1}-\cite{ref-channel-receiver-agnostic-1}.
For example, the method in \cite{ref-channel-agnostic-1} only considers the variable channel effects, assuming the same receiver during the training and testing stages.
Similarly, the authors in \cite{ref-receiver-agnostic-1} conduct the subdomain adaptation-based RFFI method on cross-receiver scenarios, but neglect the channel effects. 
Although the approach in \cite{ref-channel-receiver-agnostic-1} jointly considers these two effects, it requires extensive labeled datasets from multiple receivers for adversarial training.
Moreover, from a practical and economical perspective, a portable RFFI is hoped to be well deployed across receivers with a low price and then can be worked effectively without any additional retraining or tuning in the running time. 
As a result, designing such an RFFI method motivates our investigation in this work.

\subsection{Related Works}
In the initial phase, various works on RFFI are designed in terms of handcrafted features \cite{ref-handcraft-1}-\cite{ref-handcraft-3}.
Subsequently, with the astonishing development of artificial intelligence, the use of deep learning (DL) and convolutional neural networks (CNN) in RFFI system has received increasing attention from researchers, as these approaches can process raw signals and directly make predictions without feature engineering \cite{ref-DL-1}-\cite{ref-DL-5}.
In this paper, we systematically categorize the recent RFFI works according to the specific portability challenges that they expect to address.

\subsubsection{Channel Portability}
To solve this issue, some primary studies have been conducted with outstanding results. Specifically, the authors in \cite{ref-channel-agnostic-2} extended radio frequency-distinct native attributes (RF-DNA) fingerprinting to identify the radios under Rayleigh fading conditions.
A Nelder-Mead (N-M) simplex-based channel estimator is employed to alleviate the channel effects.
However, this method requires additional complex operations to perform channel estimation and equalization, and its performance is highly dependent on the accuracy of the estimated channel information.
The authors in \cite{ref-channel-agnostic-3} proposed a ConvMixer networks-based RFFI system, where a lightweight transfer learning algorithm is designed to retrain the classification model.
However, this method requires continuously updating its parameters to ensure adaptability to the varying channel environments, which is impractical for resource-constrained devices.
The method in \cite{ref-channel-agnostic-4} leveraged the time correlation of the channel frequency response (CFR) to resist the channel variability, where the short-time Fourier transform (STFT) algorithm is employed to construct the channel-independent spectrogram.
However, it requires different STFT symbols in the preamble to construct the spectrogram for remaining the distinct RFFs, and the phase information is neglected in their model.
In \cite{ref-channel-agnostic-1}, we proposed a spectral circular shift division based method to generate the channel-robust spectral quotient (SQ) signals, where the frequency correlation of the CFR was fully exploited.
However, the stability of handcrafted RFFs extracted from the SQ signals is compromised by the presence of outliers, resulting in a decrease in classification accuracy, especially in the low signal-to-noise (SNR) region.
Besides, more works towards this challenge can be found in the literature \cite{ref-channel-agnostic-5}-\cite{ref-channel-agnostic-7}.

\subsubsection{Receiver Portability}
There are several works that focus on this topic.
For example, the method in \cite{ref-receiver-agnostic-2} employed a known golden reference to extrapolate the differences between the RF receivers, thus removing the bias introduced by receivers.
However, the function for bais removal is obtained from a single transmitter so that it lacks generality.
In \cite{ref-receiver-agnostic-3}, the authors used a trainable transform network to calibrate the receiver effects.
However, the method also requires collecting well-labeled samples from each transmitter to update the trainable network on each newly deployed receiver.
The authors in \cite{ref-receiver-agnostic-4} proposed a weighted feature classification scheme combined with an RFFs calibration method, which can overcome the portability problem caused by receiver differentiation.
However, the calibration method only takes the carrier frequency offset (CFO) into consideration, while neglecting the effects of in-phase/quadrature (IQ) imbalance and power amplifier (PA) nonlinearity in the receiver chain.
The recent work \cite{ref-receiver-agnostic-1} jointly employed the contrastive learning and subdomain adaptation method to train a cross-receiver classifier, which achieved outstanding performance on this issue.
It should be noted that their classifier is trained with both labeled training data and unlabeled test data, while it is infeasible to touch the test data in practice.

\subsubsection{Channel and Receiver Portability}
The portability problem of multiple sources is more crucial and difficult to deal with.
To this end, the work in \cite{ref-channel-receiver-agnostic-2} proposed a multi-loss-based DL approach to project the equalized signals into a feature space, in which the transmitter fingerprints can be disentangled from the receiver fingerprints.
However, they require multiple labeled datasets collected from different receivers. 
The authors in \cite{ref-channel-receiver-agnostic-3} proposed a lightweight domain adaptation method named Tweak, which first employed the metric learning technique to train a twin neural network and then designed a distance-based decision ruler using the centroids and distances of network outputs of different devices.
When applied to a new target domain, this method only needs to calibrate the corresponding centriods and distances to improve classification performance.
However, it requires an additional well-labeled dataset of each transmitter from the target domain for its calibration task.
The authors in \cite{ref-channel-receiver-agnostic-1} conducted an adversarial training-based method to learn the receiver-independent features from the channel-independent spectrogram, which achieved significant accuracy improvements in various experiments.
However, the generated spectrogram still retains the channel effects when encountering low-end receivers (i.e., the receiver impairments cannot be neglected before dealing with the channel effects), thus a fine-tuning technique is required for the underperformed networks.
Similarly, inspired by the generative adversarial network (GAN), the authors in \cite{ref-channel-receiver-agnostic-4} proposed a two-stage supervised learning framework based RFFI method, which employed the equalized signals to extract the receiver-agnostic RFFs for device recognition.
However, this method only achieves an average correct accuracy of 68{\%} in its experiments, which is relatively poor.

\begin{figure*}[t]
\centering  
\setlength{\abovecaptionskip}{0cm}
\vspace*{-5pt}
\includegraphics[width=6.5in]{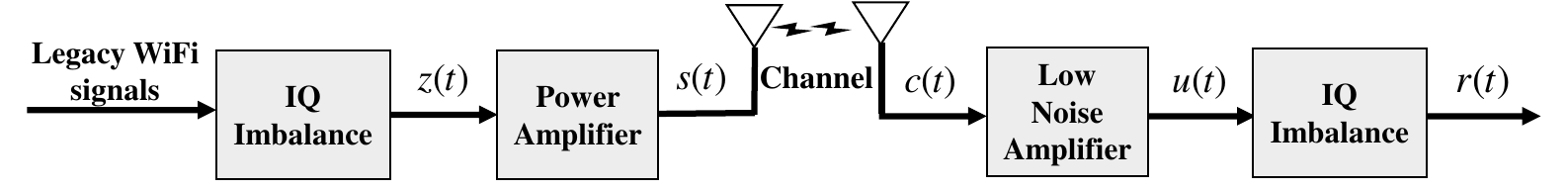}
\caption{The generation flow chart of the WiFi signal impaired with coupled RFFs.} 
\end{figure*}

\subsection{Contributions}
Motivated by solving the channel and receiver portability problems, we propose a novel two-step RFFI protocol, which exhibits robustness to channel variability and facilitates effective deployment across receivers through a straightforward calibration module.
The main contributions of our work are summarized as follows:
\begin{enumerate}
\item[$\bullet$] A channel-robust neural network for RFFI is proposed in the enrollment stage. 
By leveraging the repetitive symbols in the preamble part, we first convert the signal of interest (SOI) into a novel signal representation (Denoised Spectral Quotient, DSQ) using a channel-robust signal preprocessing module. 
Then a four-layer CNN is designed to extract the device-specific RFFs from the DSQ sequence for device identification. 
Compared to other commonly used signal representation-based approaches, the DSQCNN can effectively prevent drastic performance degradation in the unseen varying channels and provide the best classification accuracy in the high SNR regions.

\item[$\bullet$] A trainable calibration neural network (TCNN) for cross-receiver deployment is presented, which contains four dense layers and employs the augmented real-valued sequence as network input. 
We guide the TCNN to learn the nonlinear mapping relationship between receivers' effects so that it can help the well-trained DSQCNN directly deployed in a new receiver without any further processing.  
It should be noted that the data pairs of the calibration dataset used for training the TCNN are constructed by performing a phase offset (PO) based matching method on the “response” signals of different receivers.
Only one transmitter is required for data transmission during the deployment stage, suggesting that our cross-receiver method is economical and easy to implement.

\item[$\bullet$] Twelve WiFi transmitters and three receivers are configured as a simulation case study. 
On this basis, we carry out extensive simulations to investigate the channel and receiver effects on RFFI performance.
Meanwhile, we also evaluate the performance of the proposed RFFI framework, i.e., TCNN-DSQCNN, on solving the channel and receiver portability problems. 
Simulation results demonstrate that, for the cross-receiver scenarios considering unknown varying channels, the proposed TCNN-DSQCNN method achieves highly robust classification performance, boosting to more than 90$\%$ classification accuracy at SNR = 24 dB. 

\end{enumerate}

The remainder of this paper is organized as follows. 
Section II describes the considered signal model and formulates the channel and receiver portability problem.
Section III presents the proposed RFFI method.
Section IV configures the simulation settings and provides the performance results via simulations.
Section V concludes this paper.

\section{Preliminaries}
In this section, we first introduce the signal model used for simulation datasets generation, which includes the transceiver impairments and multipath fading channel effects.
The simplified block diagram of the simulation process is given in Fig. 1.
For ease of understanding, the portability problem of the joint channel and receiver is described in the following.

\subsection{Signal Model}
As provided in Fig. 2, the legacy WiFi frame is employed to generate the transmitted signal, which contains a short legacy short training field (L-STF, 8 microseconds, i.e., 8 $\mu s$), legacy long training field (L-LTF, 8 $\mu s$), legacy signal field (L-SIG, 4 $\mu s$), and data field. 
For simplicity, the modulated WiFi signal $x(t)$ is expressed as
\begin{equation}
x(t) = x_I(t) + jx_Q(t); 0\leq{t}\leq T,
\end{equation}
where $x_I(t)$ and $x_Q(t)$ represent the WiFi signal on the I and Q branches, respectively; $T$ is the time duration.

\begin{figure}[t]
\centering  
\setlength{\abovecaptionskip}{0cm}
\vspace*{-5pt}
\includegraphics[width=3.4in]{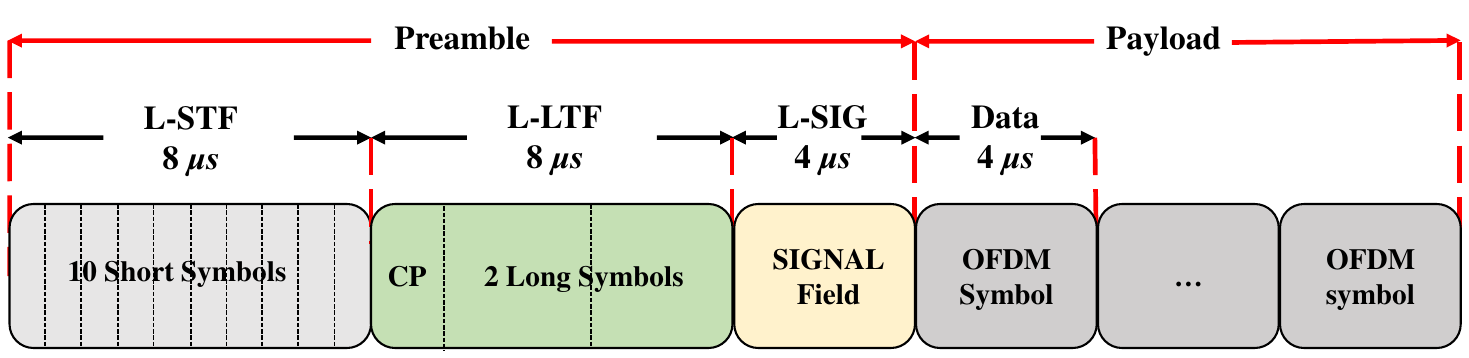}
\caption{Legacy WiFi frame structure.} 
\end{figure}

Due to the imperfect quadrature mixers, the emitted signal will be impacted by different gains and phases in the I/Q components.
Thus, the imbalanced signal $z(t)$ is denoted as
\begin{equation}
z(t) = g_I^{tx}x_I(t)e^{\frac{j\theta^{tx}}{2}} + jg_Q^{tx}x_Q(t)e^{-\frac{j\theta^{tx}}{2}},
\end{equation}
where $\theta^{tx}$ (in rad) is the phase imbalance; $g_I^{tx}$ and $g_Q^{tx}$ are the gain factors, which are computed as
\begin{equation}\label{I_im}
g_I^{tx} = 10^{0.5\frac{G^{tx}}{20}},
\end{equation}
\begin{equation}\label{Q_im}
g_Q^{tx} = 10^{-0.5\frac{G^{tx}}{20}},
\end{equation}
where $G_{tx}$ (in dB) is the gain mismatch. 

Power amplifier (PA) is a key component of the transmitter wireless communication systems, which can augment the power of the transmitted signal for a long-range transmission.
Considering the nonlinear effect of PA, the baseband distorted signal can be characterized by the Saleh model \cite{ref-Saleh} as 
\begin{equation}\label{Saleh}
s(t) = \mathscr{A}(|z(t)|)e^{j(\phi(z(t))+\mathscr{P}(|z(t)|))},
\end{equation}
where $|\cdot|$ and $\phi(\cdot)$ are the amplitude and phase operators, respectively; $\mathscr{A}(\cdot)$ and $\mathscr{P}(\cdot)$ are the functions used to describe the amplitude-modulation-to-amplitude-modulation (AM-AM) and amplitude-modulation-to-phase-modulation (AM-PM) distortions, which are given as
\begin{equation}
\mathscr{A}(|z(t)|) = \frac{\alpha_1|z(t)|}{1+\beta_1|z(t)|^2},
\end{equation}
\begin{equation}
\mathscr{P}(|z(t)|) = \frac{\alpha_2|z(t)|^2}{1+\beta_2|z(t)|^2},
\end{equation}
where $\alpha_1$, $\beta_1$, $\alpha_2$, and $\beta_2$ are the corresponding nonlinear parameters of the Saleh model.

After passing through the wireless channel, the transmitted signal will be captured at the receiver, which is denoted as
\begin{equation}
c(t) = e^{-j(2\pi \Delta f t + \Phi)}\sum_{i=1}^{I}h_{\tau_i}s(t-\tau_i) +w(t),
\end{equation}
where $\Delta f$ denotes the carrier frequency offset (CFO) between transmitter and receiver; $\Phi$ is the PO within $(-\pi, \pi)$; $h_{\tau_i}$ denotes the channel impulse of $i^{th}$ tap with delay $\tau_i$ and $I$ is the number of total delay taps.

The low noise amplifier (LNA) plays a crucial role as the first building block of the receiver chain, which commonly amplifies the received signal with good linearity.
For a low-end receiver, LNA can also introduce nonlinear distortion in the received signal.
The baseband equivalent signal modeling of nonlinear LNA is employed in this paper as it is a concise and widely adopted convention.
Thus, the baseband output of the nonlinear LNA can be described as \cite{ref-LNA-model}
\begin{equation}
u(t) = a_1c(t) + a_3c(t)|c(t)|^2,
\end{equation}
where $u(t)$ is the nonlinear LNA output; $a_1$ and $a_3$ are the nonlinear hyparameters.

After the LNA, the signal goes through a downconversion stage, thus the overall baseband received signal $r(t)$ impaired with IQ imbalance can be represented as 
\begin{equation}
r(t) = g_I^{rx}u_I(t)e^{\frac{j\theta^{rx}}{2}} + jg_Q^{rx}u_Q(t)e^{-\frac{j\theta^{rx}}{2}},
\end{equation}
where $\theta^{rx}$ is the phase mismatch in the receiver; $u_I(t)$ and $u_Q(t)$ are the I and Q components of $u(t)$, respectively; $g_I^{rx}$ and $g_I^{rx}$ can be calculated by Eq. (\ref{I_im}) and Eq. (\ref{Q_im}) with the receiver gain mismatch $G_{rx}$. This ends the signal model for simulated datasets generation.

\subsection{Problem Formulation}
\begin{figure*}[t]
\centering  
\setlength{\abovecaptionskip}{0cm}
\vspace*{-10pt}
\includegraphics[width=7in]{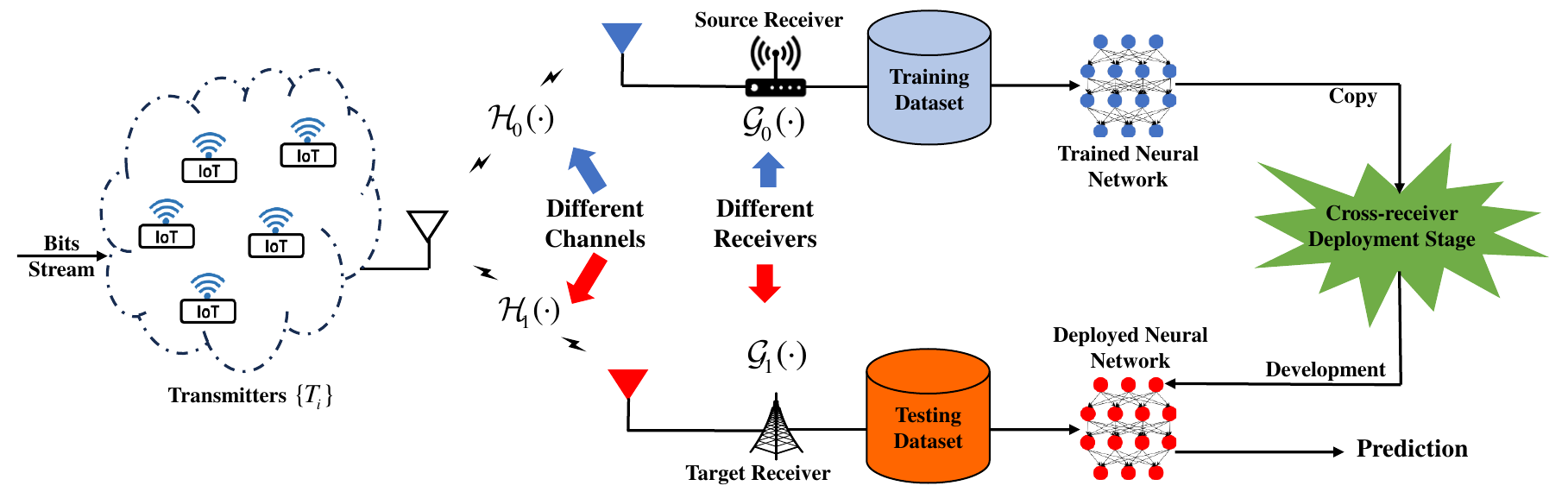}
\caption{A typical example of an RFFI system's deployment considering channel and receiver portability.} 
\end{figure*}

In this subsection, we first elaborate on the portability problem caused by the variable channels and cross-receiver deployment.
As given above, the received baseband signal can be mathematically modeled as
\begin{equation}\label{mathmodel}
r_{i}(t) = \mathcal{G}(\mathcal{H}(\mathcal{T}_i(x(t)))e^{-j(2\pi \Delta ft + \phi)}) + w(t),
\end{equation}
where $\mathcal{G}(\cdot)$ represents the hardware effects of receiver; $\mathcal{H}(\cdot)$ represents the channel effects; $\mathcal{T}_i(\cdot)$ is the unique fingerprint of the $i^{th}$ transmitter.

As shown in Fig. 3, we first employ the source receiver $\mathcal{G}_0(\cdot)$ to capture signals from different transmitters under the channel condition of $\mathcal{H}_0(\cdot)$.
Then, a total of $M$ signals will be stored as a training dataset $\{(r_m^0, l_m)\}_m^M$, where $l_m$ is the one-hot code of the transmitter used in the $m^{th}$ transmission.
Later, a well-trained neural network classifier $\mathcal{C}$ can be obtained by optimizing its parameters.

Assuming $r_k^1$ is the $k^{th}$ signal captured from the target receiver $\mathcal{G}_1(\cdot)$ with ture transmitter laber $l_k$ when the channel condition is $\mathcal{H}_1(\cdot)$.
Since both the receiver and channel effects are changed, the captured signal during the classification stage has different characteristics from that during the enrollment stage, resulting in distribution shifts.
In this case, if we directly employed the trained $\mathcal{C}$ as the deployed neural networks to classify $r_k^1$, the following prediction can be made as
\begin{equation}
\hat{l}_k = \mathcal{C}(r_k^1),
\end{equation}
where $\hat{l}_k$ is the predicted result. 
Due to the violation of the basic independent and identically distributed (i.i.d) assumption of DL, it is probable that $\hat{l}_k$ is different from $l_k$, thus degrading the performance in the cross-receiver deployment scenario.

Training the neural network classifier with the channel-independent features and the dataset including multiple receivers can be portable to the channel and receiver variations for some cases (i.e., the newly deployed receiver has similar nonlinear characteristics with one of the source receivers), but it is hard to guarantee its classification performance when deployed on a totally different receiver without further development \cite{ref-channel-receiver-agnostic-1}, \cite{ref-channel-receiver-agnostic-2}, \cite{ref-channel-receiver-agnostic-4}.
Another effective approach is to retrain or fine-tune the trained neural network during the development stage \cite{ref-channel-receiver-agnostic-3}.
The only issue with such a method is that it requires to collect enough data from all target devices through the deployed receiver, which is very luxury.
To pursue a high-accuracy classification performance in the cross-receiver deployment scenario, it is critical to design an RFFI protocol capable of solving the channel and receiver portability problem in an economical and feasible manner.

\section{The Proposed RFFI Protocol}
In this section, we introduce the workflow of the proposed RFFI protocol in terms of three stages, including the enrollment stage, cross-receiver deployment stage, and classification stage.
Then, we provide the detailed operations used in the channel-robust signal preprocessing module to generate the denoised spectral quotient (DSQ) vector.
Afterward, the DSQCNN used for the device classification is given in Subsection C.
Finally, we introduce the trainable calibration neural network (TCNN) as well as the collection process of the calibration dataset.

\subsection{System Overview}
This paper presents a novel two-stage RFFI protocol, which is robust to the channel variations and can be easily deployed on a new receiver.
The overall system is given in Fig. 4, which consists of a calibration module, a channel-robust signal preprocessing module, and a classification module.
Moreover, we can find that the development of the proposed RFFI system involves two essential stages, namely training a channel-robust DSQCNN during the enrollment stage and training a TCNN during the cross-receiver deployment stage.
After completing these training tasks, we will evaluate its performance in the classification stage.
The detailed operations in the key stages are summarized below.

\subsubsection{Operations in Enrollment Stage} 
During the enrollment stage, we first employ a source receiver to collect the signals from the transmitters.
The collected signals along with transmitter labels are served as training dataset.
Subsequently, we can suppress the channel effects through a channel-robust signal preprocessing module and then obtain the DSQ vectors.
To correctly identify the transmitters, we employ the generated DSQ vectors to train a CNN-based classifier.
Thus, the well-trained DSQCNN can be robust to the channel variations when deployed on the source receiver.

\subsubsection{Operations in Cross-receiver Deployment Stage} 
During this stage, we expect that the trained DSQCNN can be directly deployed in a new receiver.
To achieve this, we employ a calibration module cascaded with the new receiver, which enables the joint effects of them can be similar to the source receiver's effects.
Firstly, with the help of a high-end transmitter, we collect the cable-connection signals from each receiver, along with the estimated PO labels.
Then, we construct the input-output data pairs by matching the PO values, which are saved in the calibration dataset.
Finally, a simple fully-connection (dense) neural network, i.e., TCNN, can be trained using the calibration dataset. 
Thus, the overall RFFI system can be effectively worked on the cross-receiver deployment environment in the classification stage.

\begin{figure*}[t]
\centering  
\setlength{\abovecaptionskip}{0cm}
\vspace*{-5pt}
\includegraphics[width=7in]{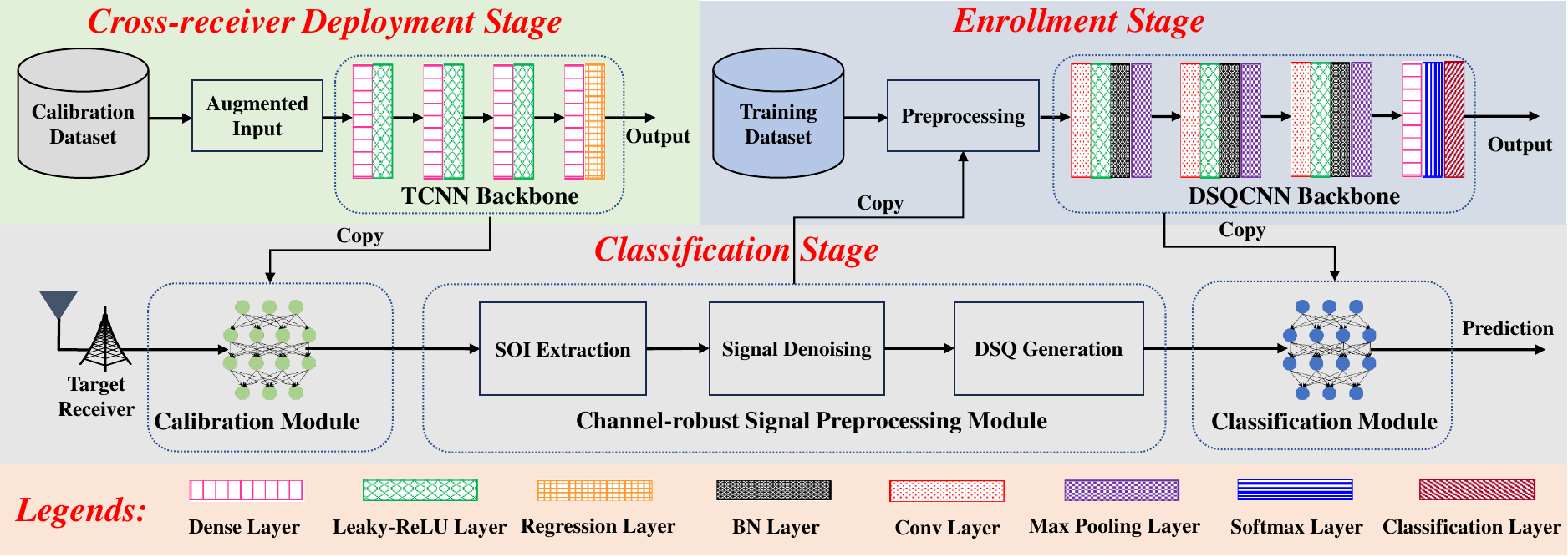}
\caption{The overall system of the proposed RFFI protocol.} 
\end{figure*}

\subsection{Channel-robust Signal Preprocessing Module} 
In the enrollment stage, our goal is to mitigate the channel effects from the received signal and then train a channel-robust CNN classifier on the source receiver $\mathcal{G}_0$.
To achieve this goal, a high-end source receiver is required to lower the receiver hardware effects.
According to Eq. (\ref{mathmodel}), the baseband signal captured by such a receiver can be simplified as 
\begin{equation}
r_{i}(t) = \mathcal{H}(\mathcal{T}_i(x(t)))e^{-j(2\pi \Delta ft + \phi)} + w(t).
\end{equation}

\subsubsection{SOI Extraction} To extract the signal of interest (SOI), the packet detection should be performed at first, which employs the autocorrelation and cross-correlation algorithm to locate the starting point of a WiFi packet \cite{ref-packet-detection}.
As referred to the previous literature \cite{ref-receiver-degrade}, CFO is unsuitable for a long-term RFFI protocol design due to the instability of the device oscillator.
Thus, we carry out the classical two-step CFO estimation algorithm to correct its effects.
Considering the preamble contains abundant RFFs information of transmitters, we employ the LTF part as the SOI.
Then, the corrected SOI in the discrete form can be approximated as
\begin{equation}
r^{SOI}(n) \approx \mathcal{H}(\mathcal{T}(x(n)))e^{-j\phi'} + w(n); 0\leq{n}\leq N_L-1,
\end{equation}
where $\phi'$ is the residual PO after CFO correction and $N_L$ is the total length of the discrete SOI.

\subsubsection{Signal Denoising} Due to the fact that the SOI contains two same OFDM symbols, a denoising method can be employed to suppress noise effects \cite{ref-denoising}. After removing its cyclic prefix (CP) of length $N_{c}$, we can obtain a denoised signal vector $\bar{\textbf{r}}^{SOI}(n) = [\bar{r}^{SOI}(0), \ldots,\bar{r}^{SOI}(n), \ldots,\bar{r}^{SOI}(N_s-1)]^T$, whose $n^{th}$ element can be expressed as
\begin{equation}
\begin{split}
\bar{r}^{SOI}(n) =r^{SOI}(n+N_{c})+r^{SOI}(&n+(N_c+N_L)/2);\\ 
&0\leq{n}\leq N_s-1,
\end{split}
\end{equation}
where $N_s = (N_L- N_c)/2$.

Next, we perform the fast Fourier transform (FFT) algorithm on the denoised signal, then a frequency-domain vector can be obtained as
\begin{equation}
R^{SOI}(k) = \sum_{n=0}^{N_s-1}\bar{r}^{SOI}(n)e^{-j2\pi nk/N_{s}}; 0\leq{n}\leq N_s-1.
\end{equation}

\subsubsection{DSQ Generation} As revealed in our previous works \cite{ref-channel-agnostic-1} and \cite{ref-channel-agnostic-5}, the channel effects can be effectively mitigated by performing the spectral circular shift division (SCSD) algorithm on the frequency-domain OFDM signal.
Considering the SOI includes both the null and active subcarriers, the SCSD algorithm should be performed in terms of the active subcarrier indices $\bm{I} = [I_0,\ldots, I_m,\ldots, I_{M}]^T$.
Thus, we first construct the division indices pairs as $\{I_n, I_{k}\}_{n=1}^{N_i}, n, k \in [0, M]$, where $|I_{n}-I_{k}| = 1$ (adjacent active subcarriers) and $N_i$ is the total number of the qualified division indices pairs. Then, the denoised spectral quotient (DSQ) signal of the division indices pair $\{I_n, I_{k}\}$ can be calculated as
\begin{equation}
\Psi(n) = R^{SOI}(I_n) /R^{SOI}(I_k),  1\leq n\leq N_i.
\end{equation}

\subsection{Classification Module}

\subsubsection{Classifier Architecture}AlexNet is a deep CNN architecture that played a pivotal role in the development of computer vision and deep learning \cite{ref-AlexNet}.
It is specially designed with eight layers, including five convolutional layers followed by three dense layers.
Inspired by this, we design a DSQCNN served as the classification module, whose backbone is shown in Fig. 4.
Unlike AlexNet, our DSQCNN is made up of four layers, namely three convolutional layers and a dense layer.
The number and size of the filters used in the convolutional layers are marked in Fig. 5.
For each convolutional layer, it is followed by batch normalization, leaky rectified linear unit (ReLU), and max pooling.
Moreover, the padding and stride in all convolutional layers are same, and they are set to 1 and 2.
Meanwhile, a dense layer is added to extract higher level non-linear combinations of the features extracted from previous layers, and the number of neurons in the dense layer is set the same as the number of the target transmitters.
Afterward, a softmax layer is used to convert the raw output scores of the dense layer into a probability vector over multiple classes.
Finally, a classification layer is adopted to output the prediction result according to the probability vector.

\subsubsection{Implementation Details}In practice, the $2\times 100$ I/Q sequence of DSQ vector is input to DSQCNN.
In order to overcome overfitting, the $\ell_2$ regularization technique is employed during the DSQCNN training, where the regularization strength is set to 0.001.
Moreover, the network parameters are trained using Adam optimizer with an initial learning rate of 0.001.
The learning rate is reduced by a factor of 0.5 for each 50 epochs.
To train our DSQCNN, we use the categorical cross-entropy as a loss function and minimize the prediction error through a back-propagation algorithm.
The total training epochs are 200 and the batch size is 320.

\begin{figure}[t]
\centering 
\setlength{\abovecaptionskip}{0cm}
\vspace*{-5pt}
\includegraphics[width=3.5in]{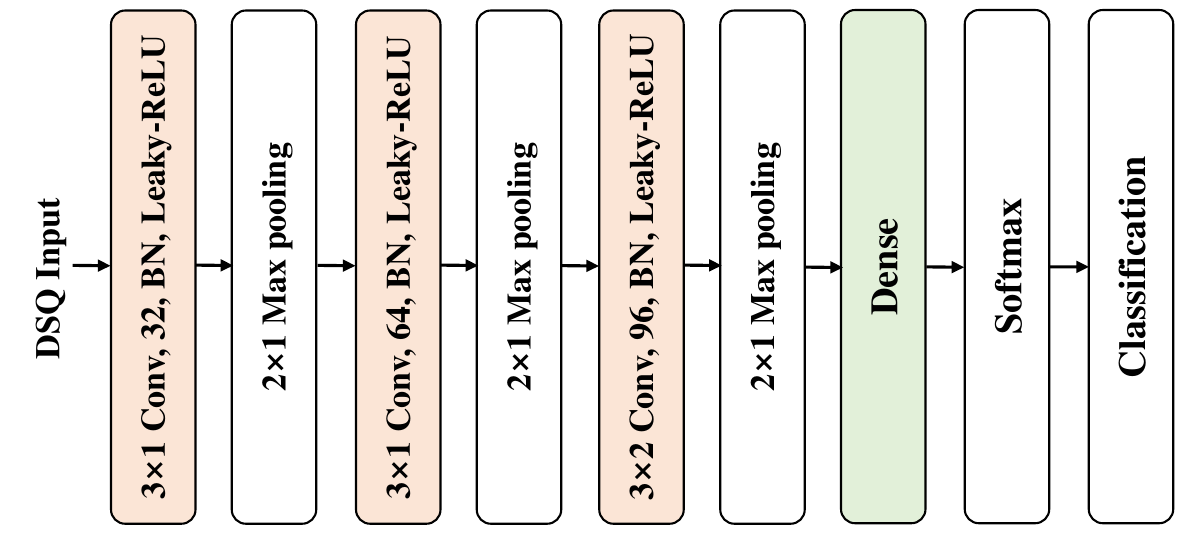}
\caption{Architecture of DSQCNN model.} 
\end{figure}

\subsection{Calibration Module}
Given a potential application of the trained classifier in a new receiver, the cross-receiver deployed classifier may result in performance degradation in the absence of appropriate countermeasures.
Thus, we propose a novel calibration technique named TCNN, which can cascade with the new receiver to solve the receiver portability issue in a feasible and low-cost manner.
To start, we first introduce the generation of the calibration dataset during the cross-receiver deployment stage.

\begin{figure}[t]
\centering  
\setlength{\abovecaptionskip}{0cm}
\vspace*{-5pt}
\includegraphics[width=3.5in]{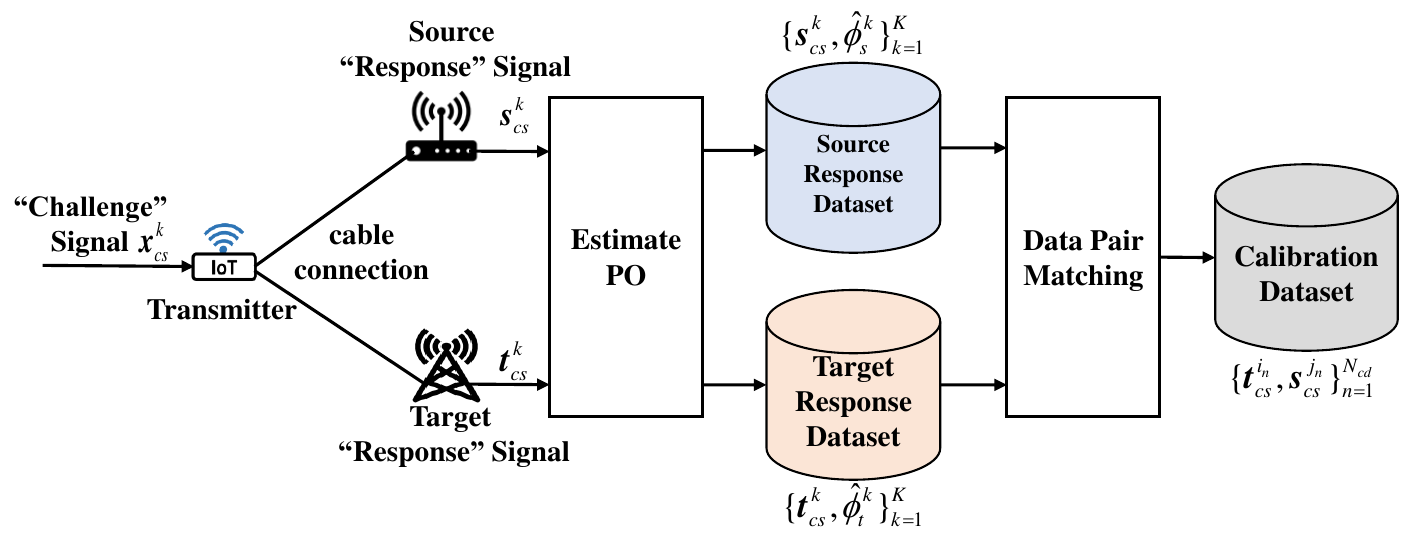}
\caption{The process of calibration dataset generation.} 
\end{figure}

\subsubsection{Calibration Dataset Generation}
Fig. 6 shows the process to generate the calibration dataset, where a high-end transmitter is required in our scheme.
First, we employ a signal frame consisting of STF and LTF as the “challenge” signals $\boldsymbol{x}_{cs}$ for transmission.
By feeding an external clock signal to both the transmitter and receiver, the CFO effects on the received signal can be almost eliminated \cite{ref-wired-CFO}.
Then, for each transmission, the “challenge” signal will be repeatedly transmitted to the source receiver and target receiver (newly deployed receiver) in the way of cable connection, respectively.
After performing frame detection using the autocorrelation algorithm, we can extract the “response” signal frames from each receiver, which can be simply modeled as
\begin{equation}
\boldsymbol{s}^k_{cs} = \mathcal{G}_0(e^{j\phi_s^k} \boldsymbol{x}_{cs}) + \boldsymbol{w}.
\end{equation}
\begin{equation}
\boldsymbol{t}^k_{cs} = \mathcal{G}_1(e^{j\phi_t^k} \boldsymbol{x}_{cs}) + \boldsymbol{w}.
\end{equation}
where $\boldsymbol{s}^k_{cs}$ and $\boldsymbol{t}^k_{cs}$ are the $k^{th}$ “response” signal frames of the source receiver and target receiver, respectively;  $\phi_s^k$ and $\phi_t^k$ are the corresponding phase offset, which can be roughly estimated as
\begin{equation}
\hat{\phi}_s^k = \frac{1}{N_{cs}}\sum_{n=1}^{N_{cs}}\phi(s^k_{cs}(n)/x_{cs}(n)),
\end{equation}
\begin{equation}
\hat{\phi}_t^k = \frac{1}{N_{cs}}\sum_{n=1}^{N_{cs}}\phi(t^k_{cs}(n)/x_{cs}(n)),
\end{equation}
where $N_{cs}$ is the length of the “challenge” signal and $\phi(\cdot)$ is the phase operator.

After repeating the above process several times, we can obtain two datasets with $K$ samples, namely source “response” dataset $\{ \boldsymbol{s}^k_{cs}, \hat{\phi}_s^k \}_{k=1}^K$ and target “response” dataset $\{ \boldsymbol{t}^k_{cs}, \hat{\phi}_t^k \}_{k=1}^K$.
By matching the PO values of the “response” signals from these two datasets, we extract the data pair with similar PO to construct the calibration dataset\footnote{It should be noted that $i_n \neq i_m$ and $j_n \neq j_m$ when $n \neq m$.} $\{ \boldsymbol{t}^{i_n}_{cs}, \boldsymbol{s}^{j_n}_{cs} \}_{n=1}^{N_{cd}}$, where $|\hat{\phi}_t^{i_n}-\hat{\phi}_s^{j_n}|\leq \varepsilon$ and $\varepsilon$ is the a predefined threshold.

Without loss of generality, we consider the differences between the input and output of a matched data pair are caused by the receiver variations.
Inspired by the digital predistortion (DPD) technique \cite{ref-DPD}, here we use a calibration module cascaded with the target receiver to reduce the receiver differences.
Unlike the DPD technique, our calibration module aims to perform the nonlinear mapping function of $\mathcal{G}_0(\mathcal{G}_1^{-1}(\cdot))$.
Since neural networks are particularly well-suited for learning nonlinear functions, our following focus is to design a simple neural network to perform the calibration task.

\subsubsection{TCNN Architecture} Inspired by the works in \cite{ref-DPD-ARVTDNN}, we design a trainable calibration neural network with an augmented real-valued input vector to learn the nonlinear mapping function. Specifically, the augmented input contains the I/Q components along with the envelope-dependent terms, which can be denoted as 
\begin{equation}
\bm{x}_{in} = [x_I,x_Q, |x|,\ldots,|x|^D]^T,
\end{equation}
where $x$ is a complex-valued sample of the matched target response signal;$D$ is the maximum order of the nonlinear term, which is set to 4 in this paper.

As shown in Fig. 7, the proposed TCNN includes four dense layers.
For the first three dense layers, each layer contains 40 neurons and employs the leaky-ReLU as an activation function.
For the last dense layer, it only has two neurons and is followed by a regression layer, which aims to predict the continuous numerical I/Q output value.

\begin{figure}[t]
\centering 
\setlength{\abovecaptionskip}{0cm}
\vspace*{-5pt}
\includegraphics[width=3.5in]{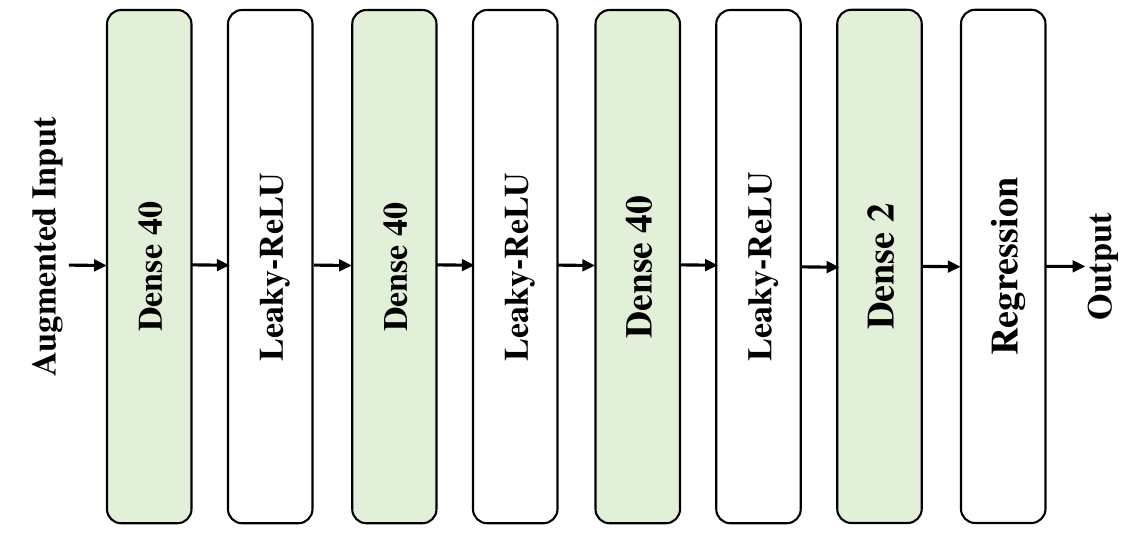}
\caption{Architecture of TCNN model.} 
\end{figure}

\subsubsection{Implementation Details}
Our TCNN is trained using Adam optimizer with a learning rate of 0.0001.
The total training epochs are 100 and the batch size is 1024. 
During the forward pass, we employ the mean square error (MSE) as the loss function, which is represented as
\begin{equation}
\text{MSE} = \frac{1}{2N_b}\sum_{n=1}^{N_b}((O_I^{p}-O_I^d)^2+(O_Q^{p}-O_Q^d)^2),
\end{equation}
where $N_b$ is the batch size; $O_I^{p}$ and $O_I^{p}$ are the predicted I/Q outputs; $O_I^{d}$ and $O_I^{d}$ are the desired I/Q outputs (i.e., the I/Q samples of the source response signal).
With the help of the back-propagation algorithm, the network hyperparameters can be iteratively updated and then the proposed TCNN can satisfy the desired nonlinear mapping function.

\section{Experimental Evaluation via Simulations}
This section first provides the simulation settings for simulated datasets generation.
Then the baseline RFFI models used for comparison and the performance metrics used for performance evaluation are introduced.
Afterward, to demonstrate the benefits of the our RFFI framework, we investigate the performance of the proposed TCNN-DSQCNN method on our simulated datasets via several well-designed experiments.
The details of our simulations are given below.

\begin{table}[t]
\renewcommand{\arraystretch}{1.5}
\vspace*{-10pt}
\caption{Impairments of Three Receivers Used in Simulations}
\vspace*{-3pt}
\centering
\begin{threeparttable}
\begin{tabular}{c|cccc|c}
\hline
\multirow{2}{*}{Receiver Code}&\multicolumn{4}{c|}{Imbalance} & \multirow{2}{*}{ LNA parameters} \\
 & \multicolumn{2}{c}{Gain (dB)} & \multicolumn{2}{c|}{Phase (degree)} & \\ 
\hline
$\mathcal{R}_0$ & \multicolumn{2}{c}{0.08} & \multicolumn{2}{c|}{2.31} & $a_3$ = -0.005 \\
\hline
$\mathcal{R}_1$ & \multicolumn{2}{c}{0.45} & \multicolumn{2}{c|}{-6.88} & $a_3$ = 0.015 \\
\hline
$\mathcal{R}_2$ & \multicolumn{2}{c}{-0.75} & \multicolumn{2}{c|}{-10.88} & $a_3$ = -0.055 \\
\hline
\end{tabular}
\begin{tablenotes}
\footnotesize
\item Notation: The parameter of $a_1$ in LNA is set to 1.
\end{tablenotes}
\end{threeparttable}
\end{table}

\subsection{Simulation Settings}
\subsubsection{Datasets Settings in the Enrollment Stage}
In this stage, we employ the legacy WiFi frame as a case study for performance evaluation, where the carrier frequency and the bandwidth are 5.765 GHz and 20 MHz, respectively.
For each frame, the length of the LTF signal is 160, where a CP with the length of 32 is included in its starting part.

To generate the simulated datasets, we configure twelve transmitters with different impairments, which are selected as follows.
The absolute gain and phase imbalances are set within the range of $[0.02,1]$ dB and $[2, 11.42]$ degree \cite{ref-RFFDataset-IQ}, respectively.
The Saleh parameters are varied within $\pm$5 of the default values $[\alpha_1 = 2.1587, \beta_1 = 1.1517, \alpha_2 = 4.0033, \beta_2 = 9.1040]$ \cite{ref-Saleh}, and the level of input back-off (IBO) from the saturation is set to 15 dB.
The CFO (both receiver and transmitter) follows the uniform distributions, whose normalized value with respect to the carrier frequency is randomly selected from the range of $[-10,10]$ ppm (parts per million).
In addition, we configure three receivers with different parameters listed in Table I, where the receiver of $\mathcal{R}_0$ is set as the source receiver and the receivers of $\mathcal{R}_1$ and $\mathcal{R}_2$ are set as the target receivers (i.e., newly deployed receiver).

Since this paper also gives a special focus on the channels, we adopt four different channel conditions to describe the multipath fading effects, whose power delay profiles (PDP) are given in Table II.
For the case of the Rician multipath fading channel, there exists a line of sight (LOS) component in its first tap and the Rician $K$-factor is set to 3.
Meanwhile, the Doppler effects are also considered and the maximum Doppler shift is set to 25 Hz in our simulations.
Besides, the channel coefficients are randomly generated for each frame according to its distribution (Rician or Rayleigh).

To sum up, a total of twelve configurations can be obtained through the combinations of channel conditions and receivers.
As a result, we generate twelve datasets used in the enrollment stage, namely $\mathcal{D}_{\mathcal{R}_0}^{\mathcal{H}_0}$, $\mathcal{D}_{\mathcal{R}_0}^{\mathcal{H}_1}$, $\mathcal{D}_{\mathcal{R}_0}^{\mathcal{H}_2}$, $\mathcal{D}_{\mathcal{R}_0}^{\mathcal{H}_3}$, $\mathcal{D}_{\mathcal{R}_1}^{\mathcal{H}_0}$, $\mathcal{D}_{\mathcal{R}_1}^{\mathcal{H}_1}$, $\mathcal{D}_{\mathcal{R}_1}^{\mathcal{H}_2}$, $\mathcal{D}_{\mathcal{R}_1}^{\mathcal{H}_3}$, $\mathcal{D}_{\mathcal{R}_2}^{\mathcal{H}_0}$, $\mathcal{D}_{\mathcal{R}_2}^{\mathcal{H}_1}$, $\mathcal{D}_{\mathcal{R}_2}^{\mathcal{H}_2}$, $\mathcal{D}_{\mathcal{R}_2}^{\mathcal{H}_3}$.
For each dataset, it contains 800 WiFi frame samples of each transmitter per signal-to-noise ratio (SNR), with the SNR range of [0, 30] dB (step is 3 dB).
In other words, there are 105600 samples in each dataset.

\begin{table}[t]
\renewcommand{\arraystretch}{1.5}
\vspace*{-10pt}
\caption{Power Delay Profiles of the Simulated Channels}
\vspace*{-3pt}
\centering
\begin{threeparttable}
\begin{tabular}{c|cccc|cc}
\hline
\multirow{2}{*}{Channel}&\multicolumn{2}{c}{\multirow{2}{*}{APG (dB)}} & \multicolumn{2}{c|}{\multirow{2}{*}{ Delay ($ns$)}} & \multicolumn{2}{c}{ Fading Type} \\
 & & & & & Rician &Rayleigh \\ 
\hline

$\mathcal{H}_{0}$ & \multicolumn{2}{c}{\multirow{2}{*}{[0, -7, -13]}} & \multicolumn{2}{c|}{\multirow{2}{*}{[0, 60, 220]}} & $\checkmark$ & \\

$\mathcal{H}_{1}$ & & & & & & $\checkmark$\\
\hline

$\mathcal{H}_{2}$ & \multicolumn{2}{c}{\multirow{2}{*}{[0, -8, -10, -15]}} & \multicolumn{2}{c|}{\multirow{2}{*}{[0, 70, 130, 270]}} & $\checkmark$ & \\

$\mathcal{H}_{3}$ & & & & & & $\checkmark$\\
\hline
\end{tabular}
\begin{tablenotes}
\footnotesize
\item Notation: APG and $ns$ are the abbreviations of average path gain and nanoseconds, respectively.
\end{tablenotes}
\end{threeparttable}
\end{table}

\subsubsection{Datasets Settings in the Deployment Stage}
As mentioned previously, we only employ the signal frame consisting of the STF and LTF as the “challenge” signal for transmission.
Since the CFO cannot be totally eliminated by the clock synchronization technique in practice, we manually add a residual CFO effect on the signals during the transmission, where the maximum residual CFO between the receiver and transmitter is set to 1000 Hz in this paper.
Meanwhile, the PO of the transmitted signal is randomly selected from the range of $[-\pi,\pi]$ for each frame.
Considering the cable-connection transmission strategy, the SNR of the received “response” signal is set to 50 dB in this case.
Besides, the maximum phase difference tolerance $\varepsilon$ is preset to 1 degree during the data pair matching process.
After performing the operations given in the Subsection D of the Section III, we can construct two simulated calibration datasets, i.e., $\mathcal{D}_{\mathcal{R}_0}^{\mathcal{R}_1}$ and $\mathcal{D}_{\mathcal{R}_0}^{\mathcal{R}_2}$.
For each dataset, it contains 800 samples of the matched “response” signal pair.

\subsection{Baseline RFFI Methods}
As learned from the previous works \cite{ref-RFFs-production} and \cite{ref-DL-4}, various signal representations can be used to support a DL-based RFFI system.
In order to demonstrate the superiority of our proposed method, thus we employ three baselines with different signal representations for performance comparison, namely raw IQ samples (RawIQ), corrected IQ samples (CIQ, after CFO correction), and FFT samples.
Moreover, we feed these power-normalized signal representations into the same CNN structure given in Subsection B of Section III, where the neuron number of their input layer is set to 128 (after removing CP).
It should be noted that we only use the simulated dataset $\mathcal{D}_{\mathcal{R}_0}^{\mathcal{H}_0}$ with 30 dB SNR to train the corresponding RFFI models.
Meanwhile, the TCNNs of different deployed receivers are trained with the datasets of $\mathcal{D}_{\mathcal{R}_0}^{\mathcal{R}_1}$ and $\mathcal{D}_{\mathcal{R}_0}^{\mathcal{R}_2}$, respectively.
For the purpose of performance evaluation and validation, these employed datasets ($\mathcal{D}_{\mathcal{R}_0}^{\mathcal{H}_0}$ of 30 dB SNR, $\mathcal{D}_{\mathcal{R}_0}^{\mathcal{R}_1}$ and $\mathcal{D}_{\mathcal{R}_0}^{\mathcal{R}_2}$) are divided into the training set, validation set, and testing set in the ratio of 5:1:2.
Moreover, the rest datasets are totally employed as the testing set in our following experiments.

\begin{figure*}[t]
	\centering  
	\vspace{-0.3cm} 
	\subfigtopskip=2pt 
	\subfigbottomskip=2pt
	\subfigcapskip=-5pt
	\subfigure[Test in $\mathcal{D}_{\mathcal{R}_0}^{\mathcal{H}_0}$]{
		\label{level.sub.1}
		\includegraphics[width=0.235\linewidth]{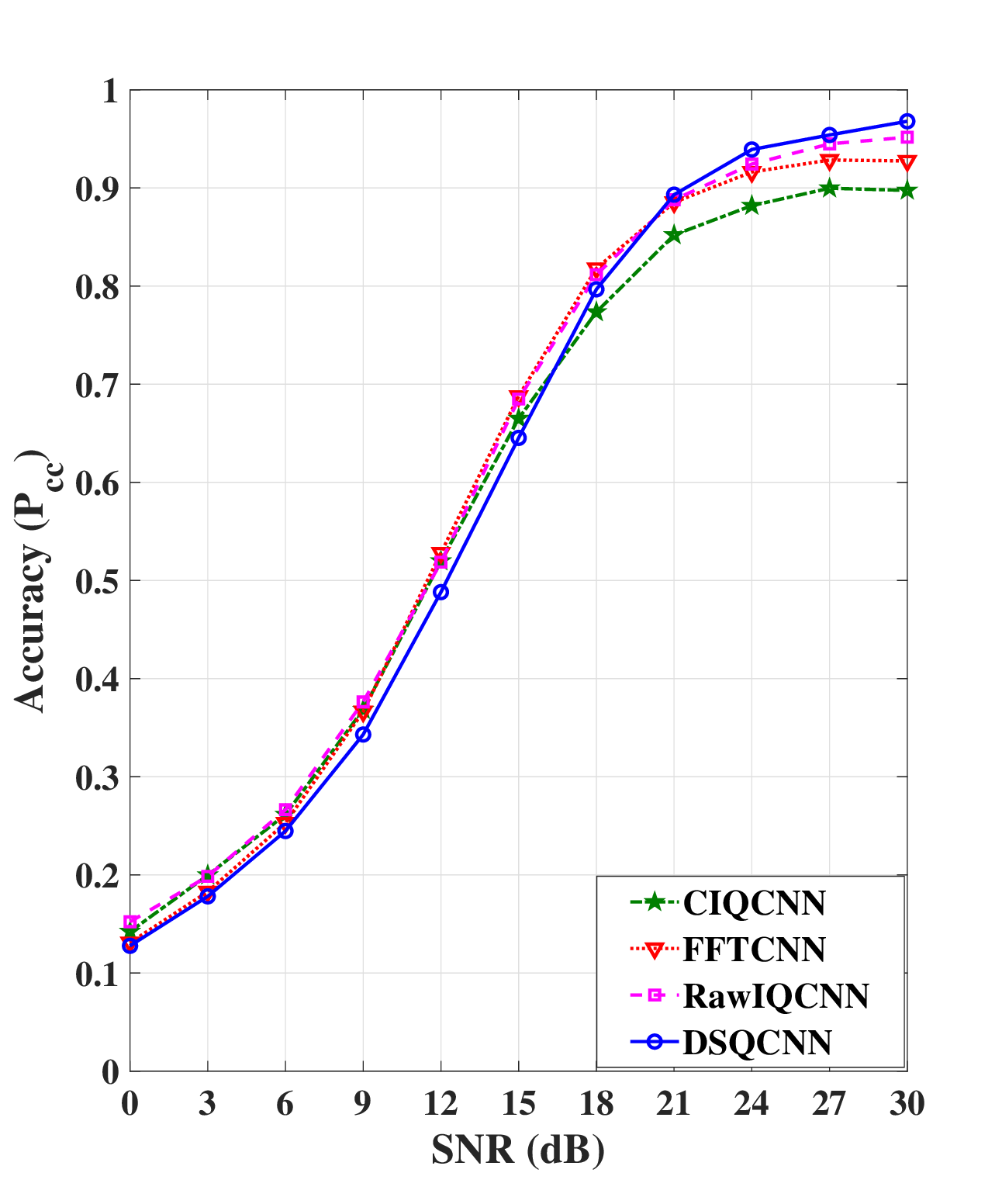}}
    \subfigure[Test in $\mathcal{D}_{\mathcal{R}_0}^{\mathcal{H}_1}$]{
		\label{level.sub.2}
		\includegraphics[width=0.235\linewidth]{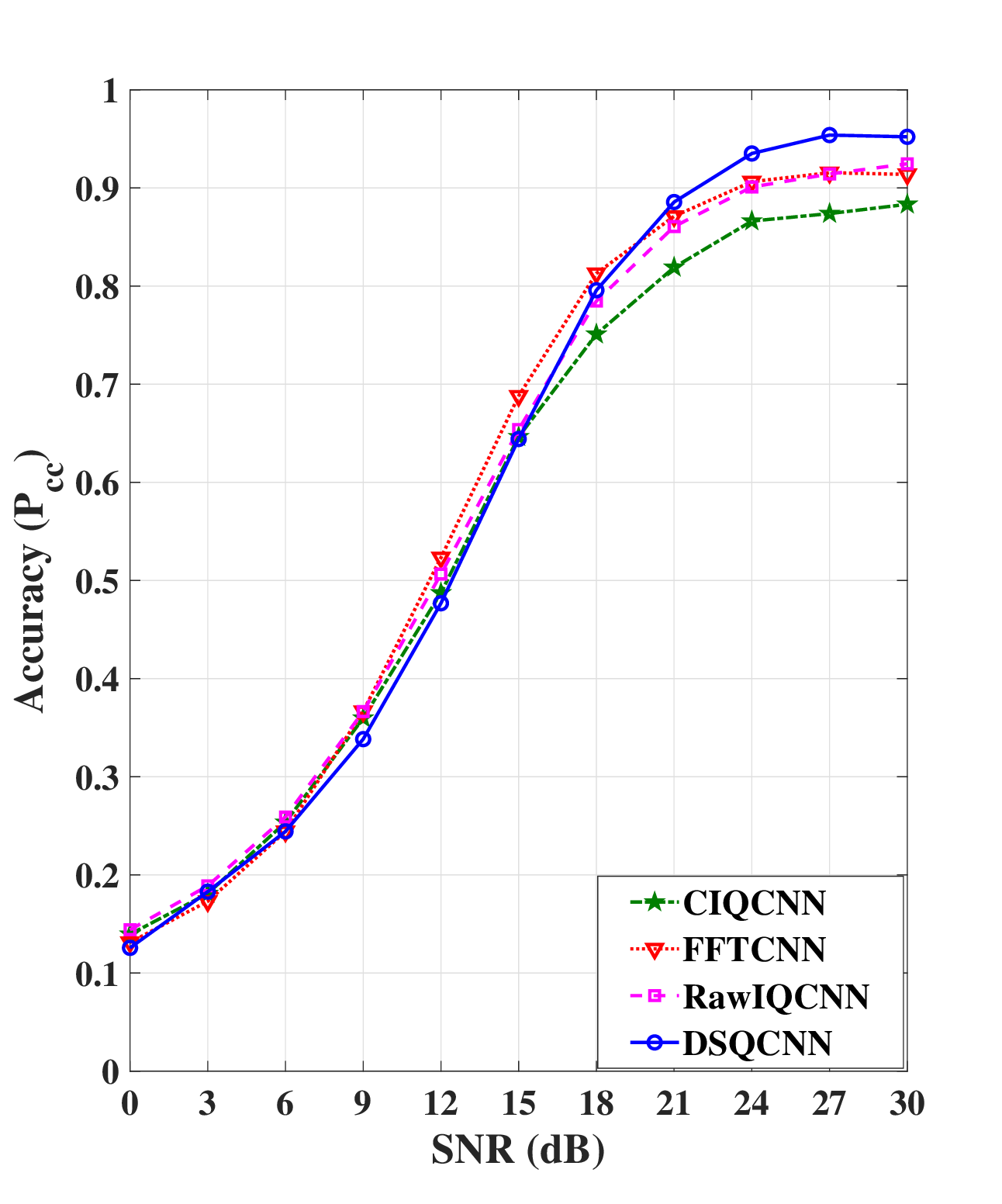}}
    \subfigure[Test in $\mathcal{D}_{\mathcal{R}_0}^{\mathcal{H}_2}$]{
		\label{level.sub.3}
		\includegraphics[width=0.235\linewidth]{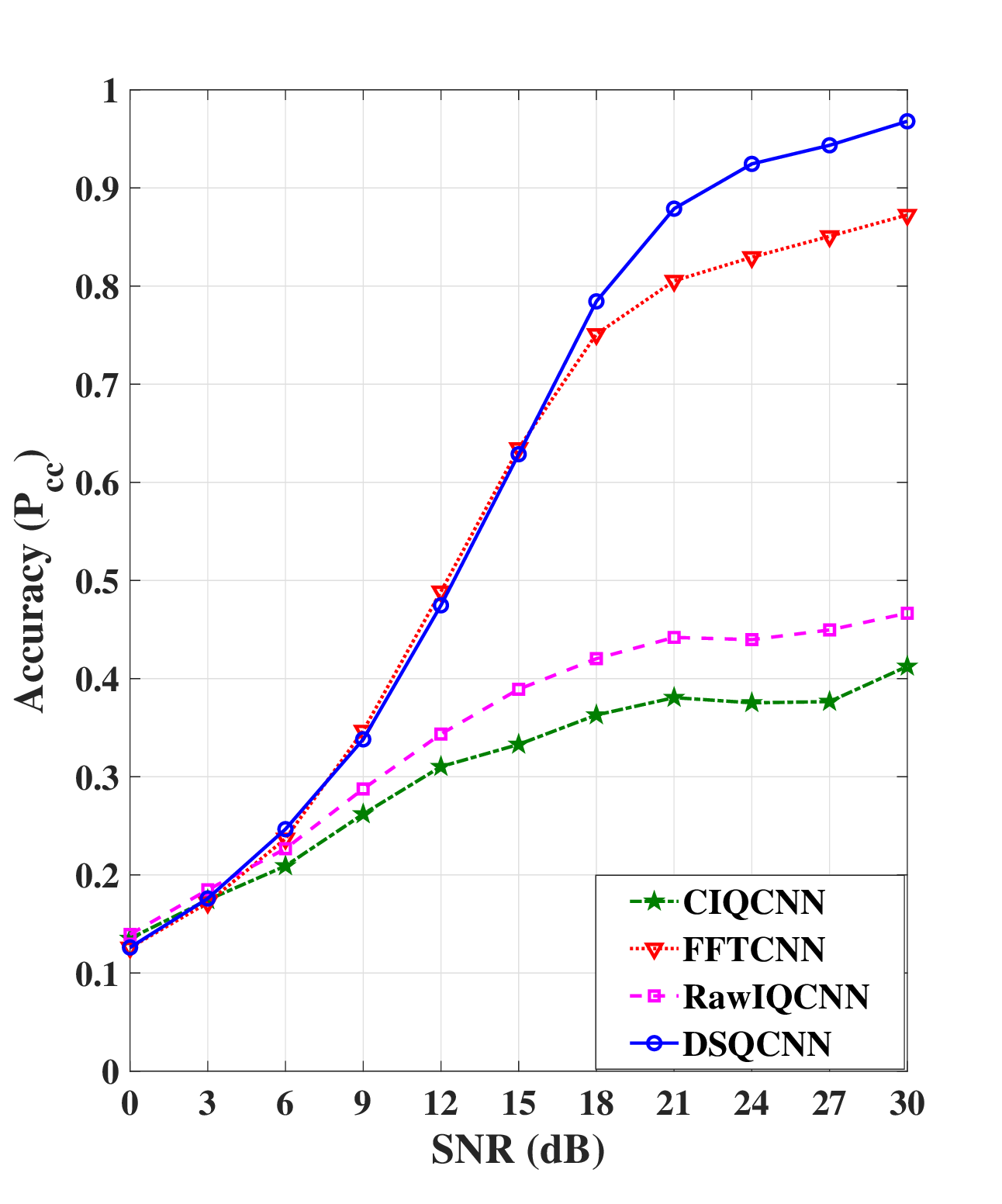}}
    \subfigure[Test in $\mathcal{D}_{\mathcal{R}_0}^{\mathcal{H}_3}$]{
		\label{level.sub.4}
		\includegraphics[width=0.235\linewidth]{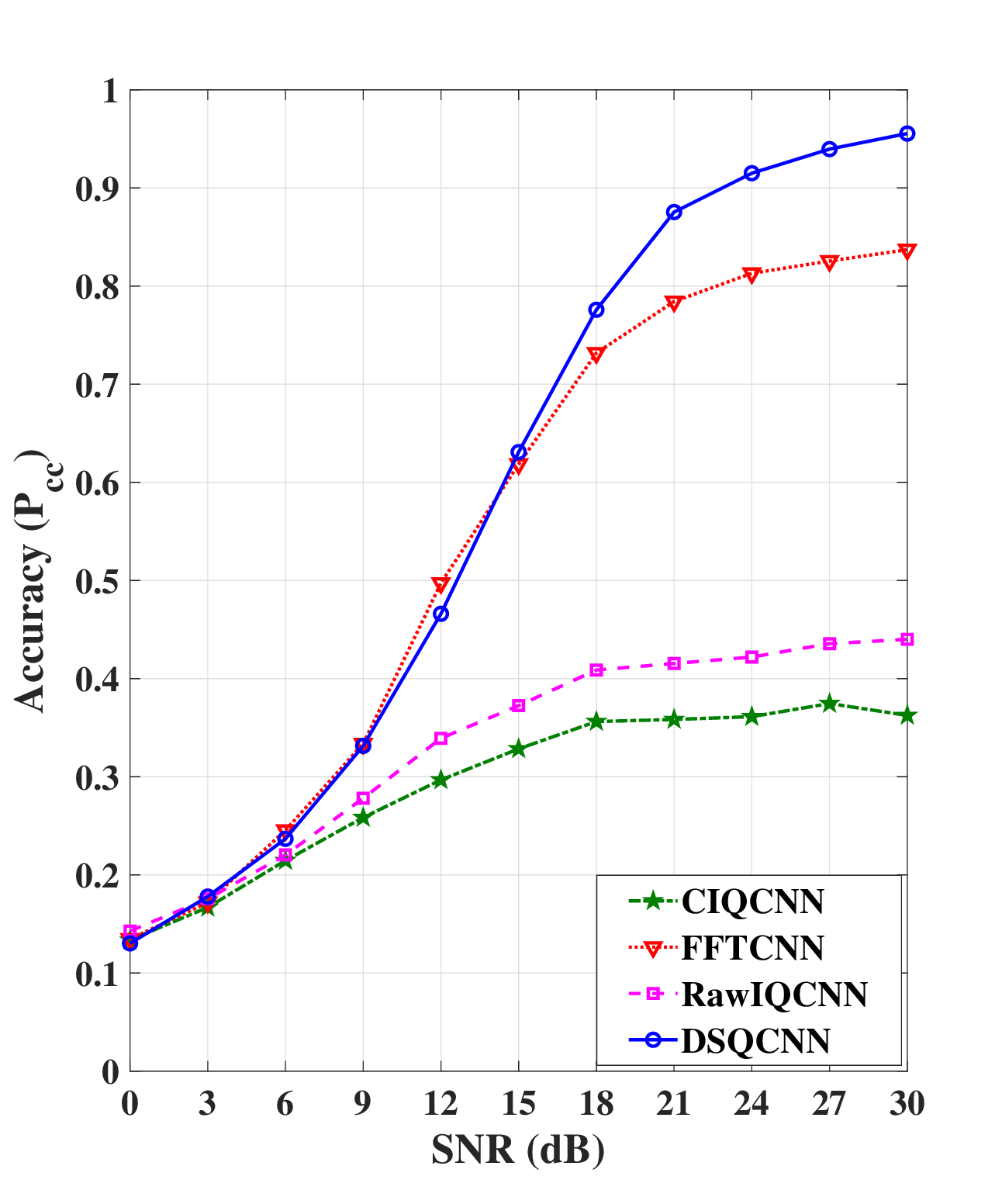}}
	\caption{Performance comparison under different channel conditions when the CNNs are trained with the dataset of $\mathcal{D}_{\mathcal{R}_0}^{\mathcal{H}_0}$.}
	\label{level}
\end{figure*}

\subsection{Performance Metrics}
Since this paper focuses on the portability problem of RFFI in a closed set, we are mainly concerned with the correct classification accuracy ($P_{cc}$) under different conditions.
Given the number of the correctly predicted samples ($T_s$) and the incorrectly predicted samples ($F_s$) at the $s^{th}$ SNR, the $P_{cc}$ is defined as
\begin{equation}
P_{cc}(s) = \frac{T_s}{T_s+F_s}.
\end{equation}

Besides, the confusion matrix is also employed here to visualize the classification performance on each device. 

For evaluating the performance of the calibration method, we employ the average normalized mean square error (NMSE) as a metric, which is calculated as
\begin{equation}
\text{NMSE} = \frac{1}{N_t}\sum_{i=1}^{N_t}10\lg(\frac{\sum_{j=1}^{N_s}|\mathcal{F}(t^{i}_{cs}(j))-s^{i}_{cs}(j)|^2}{\sum_{j=1}^{N_s}|s^{i}_{cs}(j)|^2}),
\end{equation}

where $N_t$ is the number of the matched data pairs in the testing set; $N_s$ is the length of the “response” signal; $\mathcal{F}(\cdot)$ denotes the well-trained TCNN; $t^{i}_{cs}(j)$ and $s^{i}_{cs}(j)$ are the $j^{th}$ target and source samples of the $i^{th}$ matched data pair.

\subsection{Evaluation of Channel-robust Training}
In this subsection, the impact of the variable channel conditions on RFFI systems is experimentally investigated.
To avoid the receiver-dependent RFFs, we only test the well-trained RFFI systems on the datasets generated from the same receiver, i.e., $\mathcal{D}_{\mathcal{R}_0}^{\mathcal{H}_0}$, $\mathcal{D}_{\mathcal{R}_0}^{\mathcal{H}_1}$, $\mathcal{D}_{\mathcal{R}_0}^{\mathcal{H}_2}$, $\mathcal{D}_{\mathcal{R}_0}^{\mathcal{H}_3}$.

The classification results are provided in Figs. 8, where four types of channel conditions along with different SNRs are considered.
It can observed from Fig. 8(a) that the CNNs using different signal representations are relatively effective when training and testing under the same channel condition.
Specifically, all of these methods can achieve more than 85$\%$ in accuracy when SNR is greater than 21 dB.
Moreover, according to Fig. 8(b), these methods also perform stable classification when the fading type is changed from Rician to Rayleigh.
However, as shown in Figs. 8(c) and 8(d), the performance of the CIQCNN and RawIQCNN degrades heavily by more than 45$\%$ at SNR = 30 dB when the channel's PDP is changed.
Meanwhile, the FFTCNN performs much robust than the CIQCNN and RawIQCNN in these cases, only leading to about 5$\%$-8$\%$ loss in accuracy at SNR = 30 dB.
Besides, our DSQCNN achieves stable classification performance in each dataset, where the correct classification accuracies fluctuate within 1.6$\%$ at SNR = 30 dB.
The only difference among these methods is the signal representation.
Thus, we can conclude that the DSQCNN is robust to channel variability and can achieve superior classification performance in comparison to these baselines.

\begin{table}[t]
\vspace*{-10pt}
\caption{The $P_{cc}$ Results of Different Methods When Directly \\Deployed in Cross-receiver Scenarios}
\vspace*{-3pt}
\centering
\begin{threeparttable}
\begin{tabular}{c|c|c|c}
\hline
\diagbox{Method}{$P_{cc}$}{Dataset}& $\mathcal{D}_{\mathcal{R}_0}^{\mathcal{H}_0}$ & $\mathcal{D}_{\mathcal{R}_1}^{\mathcal{H}_0}$ & $\mathcal{D}_{\mathcal{R}_2}^{\mathcal{H}_0}$  \\ 
\hline
CIQCNN &89.75\% & 63.89\% ({\color{red}25.86\% $\downarrow$}) & 50.64\%  ({\color{red}39.11\% $\downarrow$})\\ 
\hline
FFTCNN &92.75\% &66.55\% ({\color{red}26.20\% $\downarrow$})& 52.48\% ({\color{red}40.27\% $\downarrow$})  \\
\hline
RawIQCNN &95.17\% &55.41\% ({\color{red}39.76\% $\downarrow$})& 42.73\% ({\color{red}52.44\% $\downarrow$})  \\
\hline
DSQCNN &96.79\% &69.17\% ({\color{red}27.62\% $\downarrow$})& 57.89\% ({\color{red}38.90\% $\downarrow$})  \\
\hline
\end{tabular}
\begin{tablenotes}
\footnotesize
\item Notation: The red number denotes the $P_{cc}$ gap in comparison to the result achieved on $\mathcal{D}_{\mathcal{R}_0}^{\mathcal{H}_0}$.
\end{tablenotes}
\end{threeparttable}
\end{table}
\subsection{Evaluation of Receiver-portable Training}
In this subsection, we conduct the receiver effects on the performance of RFFI systems.
Firstly, we provide the $P_{cc}$ results of these methods when tested on the datasets of $\mathcal{D}_{\mathcal{R}_0}^{\mathcal{H}_0}$, $\mathcal{D}_{\mathcal{R}_1}^{\mathcal{H}_0}$, and $\mathcal{D}_{\mathcal{R}_2}^{\mathcal{H}_0}$ at SNR = 30 dB.
As shown in Table III, the change of receivers can severely compromise RFFI stability, resulting in at least a 25$\%$ accuracy drop.
The reason for the performance degradation is that the change of receivers induces the distribution shift on the received signals, which will lead to the coupled RFFs extracted on the target receiver ($\mathcal{R}_1$ or $\mathcal{R}_2$) being inconsistent with that extracted on the source receiver ($\mathcal{R}_0$), thus making the unreliable predictions on the testing data without any countermeasures.
To solve this issue, we employ a well-trained TCNN to reduce the differences in the received signals collected from the different receivers.

\begin{table}[t]
\vspace*{-10pt}
\caption{The NMSE Results With and Without TCNN \\on the Testing Samples}
\vspace*{-3pt}
\centering
\begin{threeparttable}
\begin{tabular}{c|c|c}
\hline
\diagbox{Strategy}{NMSE (dB)}{Dataset}& $\mathcal{D}_{\mathcal{R}_0}^{\mathcal{R}_1}$ & $\mathcal{D}_{\mathcal{R}_0}^{\mathcal{R}_2}$  \\ 
\hline
Without TCNN & -21.198 &-17.865 \\ 
\hline
With TCNN & -31.186 ({\color{red}9.988$\downarrow$})  &-30.766 ({\color{red}12.901 $\downarrow$}) \\
\hline
\end{tabular}
\end{threeparttable}
\end{table}

Table IV provides the NMSE results of the trained TCNN on the testing sets of $\mathcal{D}_{\mathcal{R}_0}^{\mathcal{R}_1}$ and $\mathcal{D}_{\mathcal{R}_0}^{\mathcal{R}_2}$.
It is clear that the receiver-induced differences can be effectively calibrated via our proposed TCNN module.
For instance, the NMSE results without TCNN is -21.198 dB in $\mathcal{D}_{\mathcal{R}_0}^{\mathcal{R}_1}$ and is -17.865 dB in $\mathcal{D}_{\mathcal{R}_0}^{\mathcal{R}_2}$.
Since the RFFs are subtle features extracted from the received signal, such differences can cause severe misclassification, which has been demonstrated in Table III.
After calibration, the NMSE results can be significantly reduced to -31.186 dB in $\mathcal{D}_{\mathcal{R}_0}^{\mathcal{R}_1}$ and -30.766 dB in $\mathcal{D}_{\mathcal{R}_0}^{\mathcal{R}_2}$.
Thus, by feeding the calibrated signals, the well-trained CNN can extract the coupled RFFs similar to the source receiver's and then make the correct predictions.

\begin{figure}[t]
	\centering 
	\vspace{-0.3cm} 
	\subfigtopskip=2pt 
	\subfigbottomskip=2pt 
	\subfigcapskip=-5pt 
	\subfigure[Test on the receiver $\mathcal{R}_1$]{
		\label{level.sub.1}
		\includegraphics[width=0.47\linewidth]{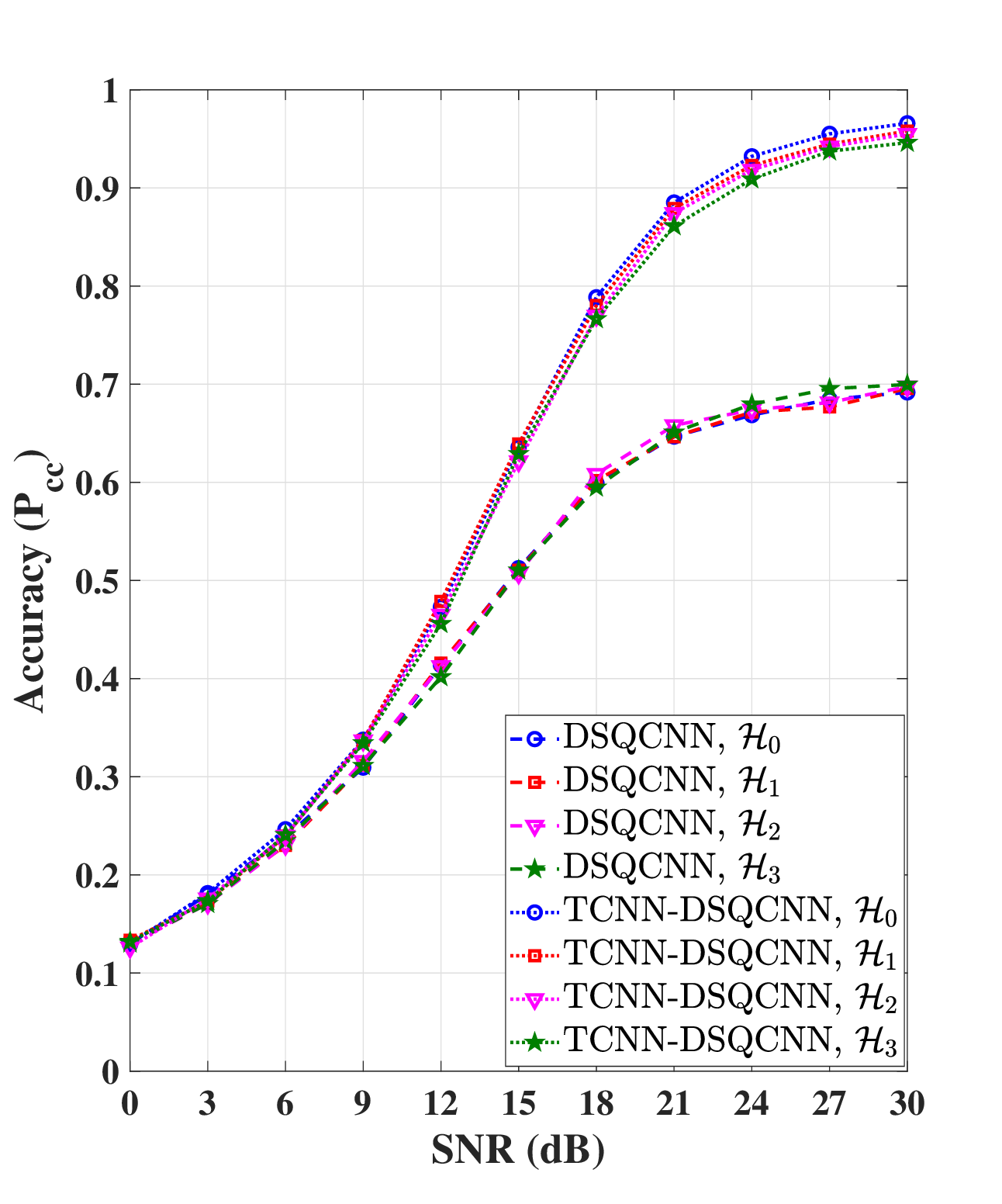}}
    \subfigure[Test on the receiver $\mathcal{R}_2$]{
		\label{level.sub.2}
		\includegraphics[width=0.47\linewidth]{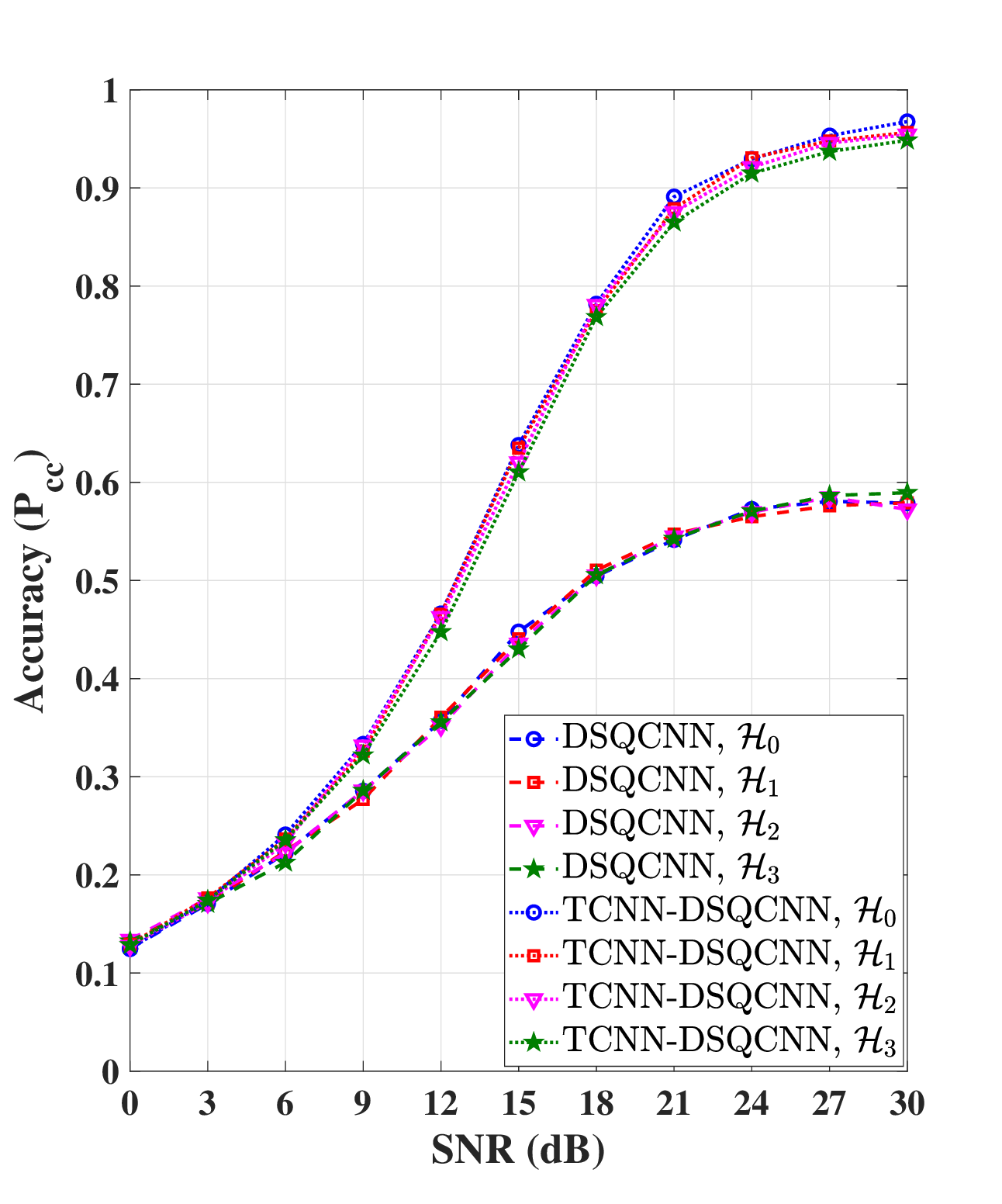}}
	\caption{Performance evaluation under different channel conditions when the DSQCNN with/without TCNN are tested in cross-receiver scenarios.}
	\label{level}
\end{figure}

Figs. 9(a) and 9(b) provide the $P_{cc}$ curves of the DSQCNN with/without TCNN when training and testing on different receivers. 
It can be appreciated from these figures that the accuracy of the TCNN-DSQCNN in all test datasets is greatly improved compared to that of solely employed the DSQCNN, where the classification accuracy can be boosted to at least 90$\%$ at SNR = 24 dB for each case.
For each channel condition, the accuracy improvement at SNR = 30 dB can reach up to about 25$\%$ on receiver $\mathcal{R}_1$ and about 35$\%$ on receiver $\mathcal{R}_2$.
Besides, in comparison to the accuracy results of DSQCNN in Fig. 8, there is almost no performance loss in all tested cross-receiver scenarios.
This can be attributed to that the calibrated signal has the similar distribution as the source received signal (used for training the DSQCNN), thus avoiding performance degradation. 
In summary, we conclude that the proposed TCNN-DSQCNN RFFI method can effectively solve the receiver and channel portability issues in the cross-receiver scenarios.

\section{Conclusions}
In this paper, we proposed a novel RFFI method named TCNN-DSQCNN to solve the channel and receiver portability problems under varying channel and cross-receiver scenarios.
We carried out systematic modeling of the hardware imperfections in the transmitter and receiver chains and then configured twelve transmitters and three receivers to generate the simulated datasets for experimental evaluations.
Through our experiments, we found that the signal representation of DSQ is suitable for training the channel-robust RFFI classifier, where the DSQCNN achieves robust and superior classification performance in comparison to the baselines using the signal representations of RawIQ, CIQ, and FFT.
As for the cross-receiver cases, the accuracy of the DSQCNN dropped more than 27.62$\%$ at SNR = 30 dB when directly deployed on a different receiver without any countermeasures.
With the aid of the TCNN module, the signal differences induced by the changes of receivers can be heavily reduced to less than -30 dB of NMSE.
Meanwhile, our TCNN-DSQCNN can obtain accuracy improvements over 25$\%$ than DSQCNN at SNR = 30 dB, which validates the effectiveness of the proposed RFFI method to the channel and receiver portability issues. 
Our future work will investigate the fusion strategies in the collaborative RFFI task.

\begin{IEEEbiography}
[{\includegraphics[width=1in,height=1.25in,clip,keepaspectratio]{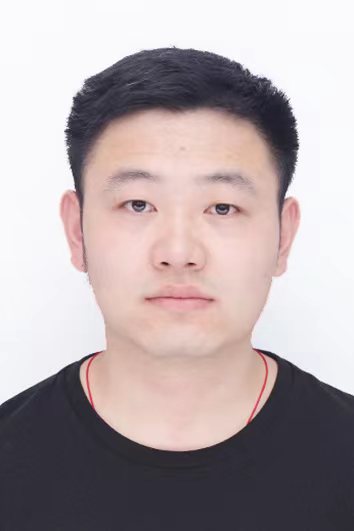}}]{Jiashuo He}
 received the M.S. degree in Communication Engineering from Xidian University, Xi’an, China, in 2021. He is the IEEE graduate student member and is currently pursuing the Ph.D. degree in Beijing University of Posts and Telecommunications, Beijing, China. His current research interests include signal processing, automatic modulation classification, and radio frequency fingerprint identification.
\end{IEEEbiography}

\vspace*{-10pt}
\begin{IEEEbiography}
[{\includegraphics[width=1in,height=1.25in,clip,keepaspectratio]{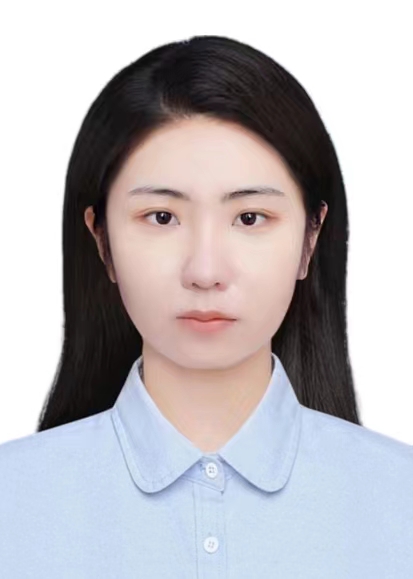}}]{Yumeng Wang}
 received the B. Eng. degree from Minzu University of China, Beijing, China in 2022.  She is currently pursuing Ph.D. degree at Beijing University of Posts and Telecommunications, Beijing, China. Her research interests focus on radio frequency fingerprint identification, positioning techniques and signal processing.
\end{IEEEbiography}

\vspace*{-10pt}
\begin{IEEEbiography}
[{\includegraphics[width=1in,height=1.25in,clip,keepaspectratio]{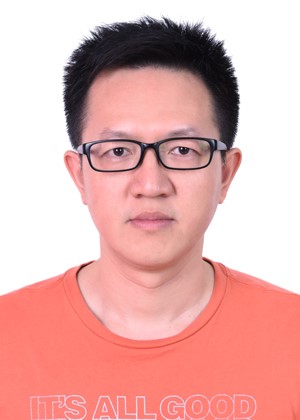}}]{Feiyang He}
is currently working at Unit 31007 of the PLA, Beijing, China.
His research directions are, electromagnetic spectrum management, intelligent spectrum sharing, and international coordination of spectrum resources.
\end{IEEEbiography}

\vspace*{-10pt}
\begin{IEEEbiography}
[{\includegraphics[width=1in,height=1.25in,clip,keepaspectratio]{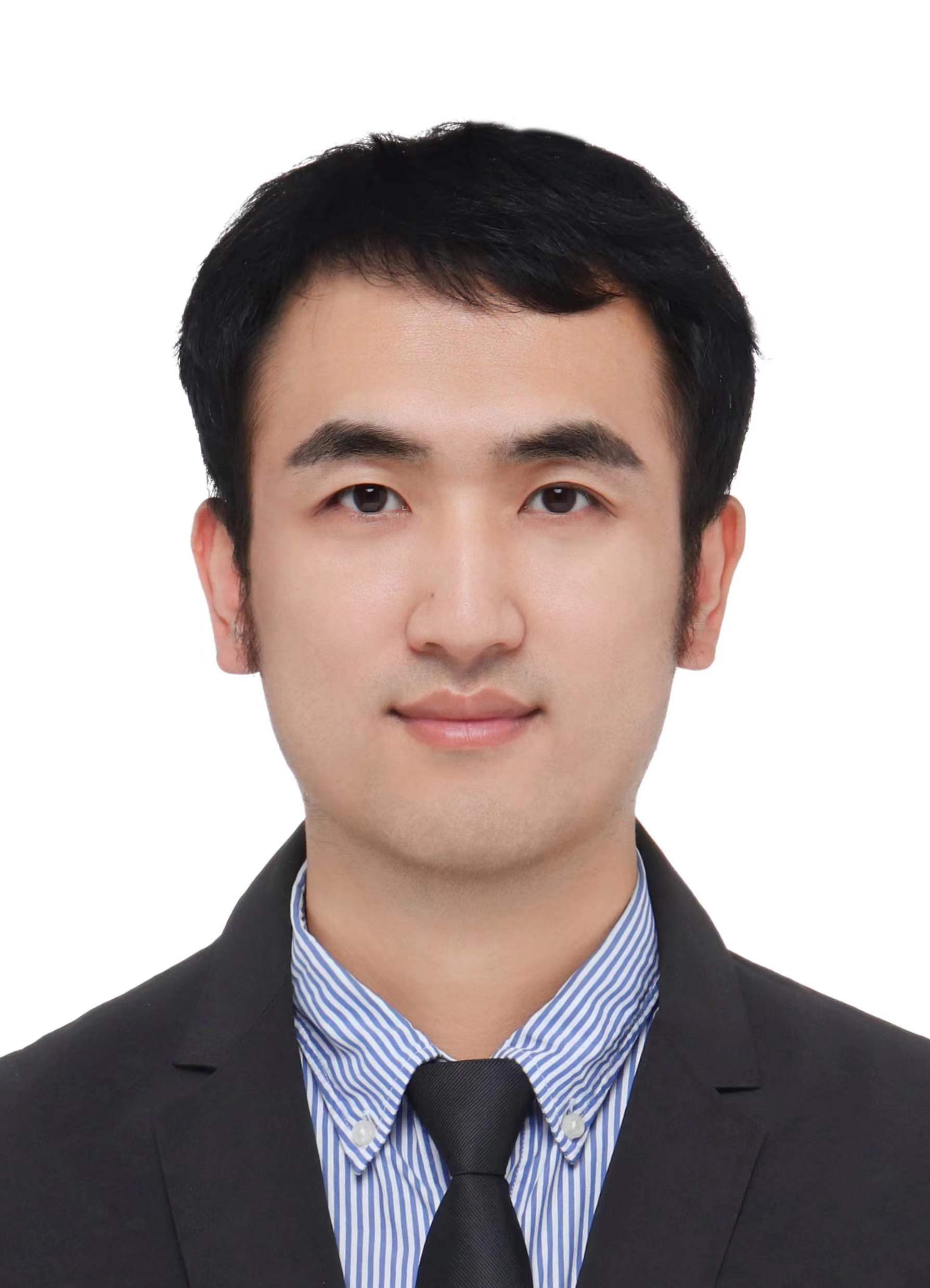}}]{Sai Huang}
 is currently working at the Department of Information and Communication Engineering as an associate professor, Beijing University of Posts and Telecommunications, and serves as the academic secretary of the Key Laboratory of Universal Wireless Communications, Ministry of Education, P.R. China.
He is the IEEE senior member and the reviewer of international journals such as IEEE Transactions on Wireless Communications, IEEE Transactions on Vehicular Technology, IEEE Wireless Communications Letters, IEEE Transactions on Cognitive Communications and Networking and International Conferences such as IEEE ICC and IEEE GLOBECOM. His research directions are machine learning assisted intelligent signal processing, statistical spectrum sensing and analysis, fast detection and depth recognition of universal wireless signals, millimeter wave signal processing and cognitive radio network.
\end{IEEEbiography}

\vspace*{-10pt}
\begin{IEEEbiography}
[{\includegraphics[width=1in,height=1.25in,clip,keepaspectratio]{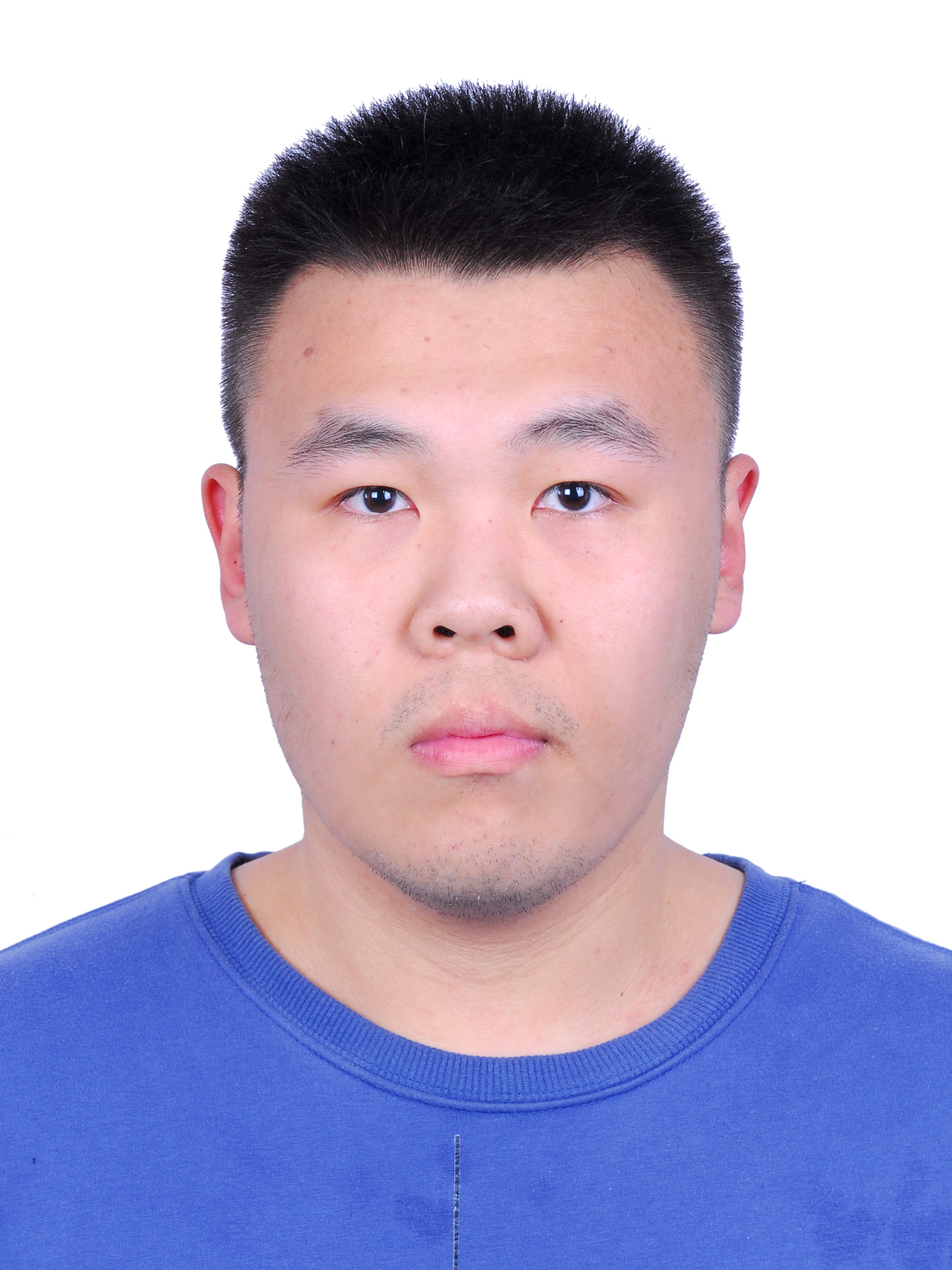}}]{Yiheng Liu}
 received the B.E. degree from Beijing University of Posts and Telecommunications in 2022. He is currently pursuing the M.S. degree at BUPT. His research interests focus on physical layer decoding, dexterous interference of 5G-NR systems,and intelligent signal classification.
\end{IEEEbiography}

\vspace*{-10pt}
\begin{IEEEbiography}
[{\includegraphics[width=1in,height=1.25in,clip,keepaspectratio]{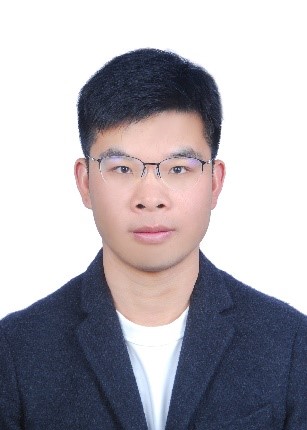}}]{Shuo Chang}
 received the B.S. degree in communication engineering from Shenyang Jianzhu University, Shenyang, China, in 2015. And he received Ph.D. degree from the Beijing University of Posts and Telecommunications (BUPT), China, in 2020. He has made a post doc in school of computer science in BUPT. He is currently a lecturer of School of Cyberspace Security of BUPT. His research interests include signal processing, visual object tracking, visual detection, and sensor fusion. 
\end{IEEEbiography}

\vspace*{-10pt}
\begin{IEEEbiography}
[{\includegraphics[width=1in,height=1.25in,clip,keepaspectratio]{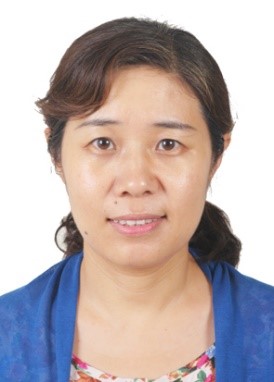}}]{Zhiyong Feng}
is a senior member of IEEE and a full professor. She is the director of the Key Laboratory of Universal Wireless Communications, Ministry of Education. She holds B.S., M.S., and Ph.D. degrees in Information and Communication Engineering from Beijing University of Posts and Telecommunications (BUPT), Beijing, China. She is a technical advisor of NGMN, the editor of IET Communications, and KSII Transactions on Internet and Information Systems, the reviewer of IEEE TWC, IEEE TVT, and IEEE JSAC. She is active in ITU-R, IEEE, ETSI and CCSA standards. Her main research interests include wireless network architecture design and radio resource management in 5th generation mobile networks (5G), spectrum sensing and dynamic spectrum management in cognitive wireless networks, universal signal detection and identification, and network information theory.
\end{IEEEbiography}

\end{document}